\documentclass[aps,prd,floatfix,showpacs,10pt,twocolumn,nofootinbib]{revtex4-1}
\usepackage{amsmath}
\usepackage{amsfonts}
\usepackage{amssymb}
\usepackage{amsthm}
\usepackage{mathrsfs}
\usepackage{multirow}
\usepackage{color}
\usepackage{cancel}
\usepackage{array}
\usepackage{tikz}
\usetikzlibrary{calc}
\usetikzlibrary{decorations}
\usetikzlibrary{decorations.fractals}
\usetikzlibrary{decorations.pathreplacing}
\usetikzlibrary{decorations.pathmorphing}
\usetikzlibrary{decorations.markings}
\usetikzlibrary{positioning,matrix}
\usetikzlibrary{shapes.arrows}
\usetikzlibrary{shapes.symbols}
\usetikzlibrary{shapes.misc}
\usetikzlibrary{shapes.geometric}
\usetikzlibrary{through}
\usetikzlibrary{mindmap,backgrounds}
\usetikzlibrary{fit}
\usetikzlibrary{arrows,positioning}
\usetikzlibrary{external}
\usepackage[colorlinks, urlcolor=black, citecolor=blue, link
color=black]{hyperref}

\def\nn    {\nonumber}

\def\beq{\begin{equation}}
\def\eeq{\end{equation}}
\def\beqa{\begin{eqnarray}}
\def\eeqa{\end{eqnarray}}

\allowdisplaybreaks

\begin{document}

\title{Search for $tZ'$ associated production induced by $tcZ'$ couplings at the LHC}

\author{Wei-Shu Hou}
\affiliation{Department of Physics, National Taiwan University, Taipei 10617, Taiwan}
\author{Masaya Kohda}
\affiliation{Department of Physics, National Taiwan University, Taipei 10617, Taiwan}
\author{Tanmoy Modak}
\affiliation{Department of Physics, National Taiwan University, Taipei 10617, Taiwan}
%

\vspace{.2cm}

\begin{abstract}
The $P'_5$ and $R_K$ anomalies, recently observed by the LHCb collaboration in 
$B \to K^{(*)}$ transitions, may indicate the existence of a new $Z'$ boson, 
which may arise from gauged $L_\mu - L_\tau$ symmetry. 
Flavor-changing neutral current $Z'$ couplings, such as $tcZ'$, can be induced by the 
presence of extra vector-like quarks. In this paper we study the LHC signatures of 
the induced right-handed $tcZ'$ coupling that is inspired by, but not directly linked to, 
the $B \to K^{(*)}$ anomalies. The specific processes studied are $cg \to tZ'$ and 
its conjugate process each followed by $Z'\to\mu^+\mu^-$. 
By constructing an effective theory for the $tcZ'$ coupling, we first explore in a
model-independent way the discovery potential of such a $Z'$ at the 14 TeV LHC 
with 300 and 3000 fb$^{-1}$ integrated luminosities.
We then reinterpret the model-independent results within the gauged $L_\mu - L_\tau$ model.
In connection with $tcZ'$, the model also implies the existence of a flavor-conserving $ccZ'$ 
coupling, which can drive the $c \bar c \to Z' \to \mu^+\mu^-$ process.
Our study shows that existing LHC results for dimuon resonance searches already 
constrain the $ccZ'$ coupling, and that the $Z'$ can be discovered in either or both of 
the $cg \to tZ'$ and $c \bar c \to Z'$ processes. We further discuss the sensitivity to 
the left-handed $tcZ'$ coupling and find that the coupling values favored by the 
$B \to K^{(*)}$ anomalies lie slightly below the LHC discovery reach even with
3000 fb$^{-1}$.
\end{abstract}

\maketitle
\section{Introduction}

Recent measurements performed by the LHCb 
experiment \cite{Aaij:2013qta,Aaij:2015oid,Aaij:2014ora} exhibit anomalous 
$B\to K^{(*)}$ transitions.
One is the measurement \cite{Aaij:2013qta,Aaij:2015oid} of angular observables for the 
$B^{0} \to K^{*0} \mu^+ \mu^-$ decay, which shows a discrepancy from the Standard Model
(SM) prediction at 3.4$\sigma$ level, mainly driven by the $P_5'$ observable.
In another measurement \cite{Aaij:2014ora} of $B^{+}\to K^{+}\ell^{+}\ell^{-}$ decays 
($\ell =e$ or $\mu$), LHCb found a further hint for lepton flavor universality violation, 
namely a 2.6$\sigma$ deviation of the observable 
$R_K \equiv\mathcal{B}(B^{+}\to K^{+}\mu^+\mu^-)/\mathcal{B}(B^{+}\to K^{+}e^+e^-)$ 
from its SM value.
These LHCb results are supported by a recent Belle analysis \cite{Wehle:2016yoi}, 
where the angular observables were separately measured for the muon and electron modes
of $B \to K^{*} \ell^+ \ell^-$ decays, and the muonic $P_5'$ was found to show
the largest discrepancy (at 2.6$\sigma$ level) from the SM prediction.
Although these anomalies can well be due to statistical fluctuations and/or hadronic
uncertainties, it is interesting to investigate whether they can be attributed to 
physics beyond the SM (BSM). 
Model-independent analyses by various groups have found that a BSM contribution
to the Wilson coefficient $C^{\mu}_9$, associated with the effective operator 
$O_9^\mu = (\bar s_L\gamma^\alpha b_L)(\bar\mu \gamma_\alpha \mu)$,
can explain both the $P_5'$ ~\cite{Descotes-Genon:2013wba,Altmannshofer:2013foa,Beaujean:2013soa,Horgan:2013pva,Hurth:2013ssa}
and $R_K$~\cite{Alonso:2014csa,Hiller:2014yaa,Ghosh:2014awa} anomalies, 
by a similar amount in BSM effect~\cite{Hurth:2014vma,Altmannshofer:2014rta}.

Given the $B\to K^{(*)}\ell^+\ell^-$ data suggest BSM effects in the muon modes 
rather than the electron modes, an interesting BSM candidate is a new gauge boson
$Z'$ of the gauged $L_\mu-L_\tau$ symmetry~\cite{He:1990pn,Foot:1990mn}, 
the difference between the muon and tau numbers.
The $Z'$ boson couples to the muon but not to the electron.
In Ref.~\cite{Altmannshofer:2014cfa}, an extension of the gauged $L_\mu-L_\tau$
symmetry was constructed for sake of introducing flavor-changing neutral current (FCNC)
$Z'$ couplings to the quark sector.
In the model, the SM quarks mix with new vector-like quarks, that are charged under the 
new gauge symmetry, leading to effective FCNC couplings of $Z'$ with SM quarks. 
Among these, the left-handed (LH) $bsZ'$ coupling gives rise to $C^{\mu}_9$.
The model provides a viable explanation for both the $P_5'$ and $R_K$ anomalies.

The gauged $L_\mu-L_\tau$ model is, however, just one possibility among many options 
for a UV theory. Hence, the model should be cross-checked by other ways, in particular, 
by direct searches at colliders. LHC phenomenology within the minimal version of the gauged 
$L_\mu-L_\tau$ model has been studied in Refs.~\cite{Altmannshofer:2014cfa,Harigaya:2013twa,Altmannshofer:2014pba,delAguila:2014soa,Elahi:2015vzh}, 
where $Z'$ is searched in $Z\to \mu^+\mu^- Z'(\to \mu^+\mu^-)$.
The search is sensitive to $Z'$ lighter than the $Z$ boson and can probe the new gauge 
coupling $g'$ as well as the $Z'$ mass $m_{Z'}$.
On the other hand, the extended model \cite{Altmannshofer:2014cfa} gives effective 
$Z'$ couplings to SM quarks, and these couplings could offer new ways to produce
the $Z'$ boson at colliders. In particular, the model predicts the existence of 
not only a LH $tcZ'$ coupling that is directly related to the LH $bsZ'$ coupling by 
SU(2)$_L$ gauge symmetry, but also a right-handed (RH) $tcZ'$ coupling.
Refs.~\cite{Altmannshofer:2014cfa,Fuyuto:2015gmk} have studied $t\to c Z'$ decay
induced by these $tcZ'$ couplings.
This decay can be searched for in the huge number of $t\bar t$ events at the LHC;
however, it becomes kinematically forbidden if the $Z'$ mass is greater than the mass 
difference between the top and charm quarks, i.e. for $m_{Z'} > m_t-m_c$.\footnote{
 For $m_{Z'}>m_t+m_c$, $Z'\to tc$ \cite{Arhrib:2006sg,Cakir:2010rs,Aranda:2010cy}
 may happen, but its branching ratio is highly suppressed due to mixings between 
 the heavy vector-like and SM quarks, in addition to rather low $Z'$ production 
 cross sections in the model we consider.
}

In this paper we consider another unique production mechanism of the $Z'$ boson via 
the $tcZ'$ couplings, namely, $cg\to tZ'$. To be specific, we study the following processes 
at the 14 TeV LHC: $p p \to t Z'$ (hereafter denoted as the $t Z'$ process) and its conjugate
$ p p \to \bar{t} Z'$ (denoted as $\bar{t}Z'$) process, each followed by 
$Z' \to \mu^+ \mu^-$ and $t\to bW^+(\to \ell^+\nu_\ell)$ (or its conjugate).
A model-independent study of such $tcZ'$-induced processes at the LHC has been 
performed in Ref.~\cite{Gupta:2010wt}.\footnote{
 A $tuZ'$-induced process $ug\to tZ'$ has also been studied with $Z'$ decays to quarks in
 Ref.~\cite{Gresham:2011dg} ($Z'\to tj$) and \cite{Ng:2011jv} ($Z'\to b\bar b$).
}
We improve the treatment of SM background processes 
by including the ones missed in the previous study
and find that the $\bar{t}Z'$ process is better
suited for discovery than $tZ'$ due to lower background.
Combining the two signal processes (also referred to as the $tZ'$ process collectively
if there is no confusion),
we present first the model-independent discovery potential of 
the $t Z'$ process, aiming for the high-luminosity LHC (HL-LHC).
In detailing our collider analysis, we choose two representative $Z'$ mass values: 
just below (150 GeV) and above (200 GeV) the top-quark mass. 
We then extend the latter case to $Z'$ masses up to 700 GeV,
and reinterpret the model-independent results for RH $tcZ'$ coupling within 
the gauged $L_\mu-L_\tau$ model~\cite{Altmannshofer:2014cfa}. 
It turns out that the LH $tcZ'$ coupling implied by the $B\to K^{(*)}$ anomalies 
is rather small, and lies slightly beyond the discovery reach of the LHC even with 3000 fb$^{-1}$ data. 
Therefore, we mainly focus on the RH $tcZ'$ coupling, which is 
hardly probed by B physics. Yet our results can be easily translated into the case of
the LH $tcZ'$ coupling.

The model implies a flavor-conserving effective $ccZ'$
coupling along with $tcZ'$, while the effective $Z'$ couplings containing 
the up quark, i.e. $uuZ'$, $cuZ'$ and $tuZ'$, are suppressed by $D$ meson constraints.
The $ccZ'$ coupling offers another production channel for $Z'$ at the LHC, i.e.
$c\bar c \to Z' \to \mu^+ \mu^-$ (hereafter denoted as the dimuon process).
Analogous to the $t Z'$ case, we first perform a model-independent study,
which is then reinterpreted within the gauged $L_\mu-L_\tau$ model.
We find that the $Z'$ can be discovered in either or both of the $t Z'$ and 
dimuon processes.
We show that the dimuon process has a better chance for discovery in most of the 
model parameter space, while simultaneously measuring the $t Z'$ process can 
confirm the flavor structure of the $Z'$ model.

The paper is organized as follows. In Sec.~\ref{formalism}, we briefly introduce 
the gauged $L_\mu-L_\tau$ model of Ref.~\cite{Altmannshofer:2014cfa} 
and give the effective Lagrangian for $tcZ'$ and $ccZ'$ couplings. 
We detail our collider analysis in Sec.~\ref{direct}, which is divided into two subsections:
the $\bar{t}Z'$ and $tZ'$ processes induced by $tcZ'$ coupling in Sec.~\ref{zptc}, 
and the dimuon process induced by $ccZ'$ coupling in Sec.~\ref{dimuonc}.
In Sec.~\ref{zptc}, we also utilize an existing LHC data~\cite{Sirunyan:2017kkr} 
to illustrate its implication for $tcZ'$ coupling.
Three subsections are assigned to Sec.~\ref{const}. 
In Sec.~\ref{effcoupling} we present the model-independent discovery reaches 
for RH $tcZ'$ and $ccZ'$ couplings at the HL-LHC. 
In Sec.~\ref{glmutau}, we reinterpret the model-independent 
results within the gauged $L_\mu-L_\tau$ model. 
In Sec.~\ref{senlh} we discuss collider sensitivities to the LH $tcZ'$ coupling, 
which is directly linked to the $B\to K^{(*)}$ anomalies.
We summarize and offer further discussions in Sec.~\ref{summary}.

\section{Model}\label{formalism}

Let us briefly introduce the gauged $L_\mu-L_\tau$ model of Ref.~\cite{Altmannshofer:2014cfa},
where a new U(1)$'$ gauge group associated with $L_\mu-L_\tau$ symmetry is introduced.
The gauge and Higgs sectors of the U(1)$'$ consist of the gauge field $Z'$ and 
the SM gauge singlet scalar field $\Phi$, which carries unit charge under the U(1)$'$. 
The $\Phi$ field acquires a nonzero vacuum expectation value (VEV)
$\langle \Phi \rangle = v_\Phi/\sqrt{2}$, which spontaneously breaks the U(1)$'$ 
and gives mass to $Z'$, $m_{Z'}=g' v_\Phi$. 
In the minimal model, the $Z'$ couples to the SM fermions through 
\begin{align}
 \mathcal{L}\supset -g'\left( \bar{\mu}\gamma_\alpha \mu+\bar{\nu}_{\mu L}
 \gamma_\alpha \nu_{\mu L} -\bar{\tau}\gamma_\alpha \tau-\bar{\nu}_{\tau L}
 \gamma_\alpha \nu_{\tau L} \right)Z'^{\alpha}~\label{lepton}.
\end{align}

In Ref.~\cite{Altmannshofer:2014cfa}, an extended model was constructed by the addition 
of vector-like quarks $Q_L=(U_L, ~D_L)$, $U_R$, $D_R$ and 
their chiral partners $\tilde{Q}_R=(\tilde{U}_R, ~\tilde{D}_R)$, $\tilde{U}_L$, $\tilde{D}_L$. 
The vector-like quarks carry $+1$ U(1)$'$ charge for $Q\equiv Q_L+\tilde{Q}_R$, 
and $-1$ for $U\equiv U_R+\tilde{U}_L$ and $D\equiv D_R+\tilde{D}_L$,
with gauge invariant mass terms given by
\begin{align}
 -\mathcal{L}_{\rm mass}=m_Q \bar{Q}Q +m_U \bar{U}U +m_D \bar{D}D.
\end{align}
The vector-like quarks mix with SM quarks via Yukawa interactions given by
\begin{align}
 -\mathcal{L}_{\rm mix}
 &=\Phi\sum_{i=1}^3 \left( \bar{\tilde{U}}_{R} {Y_{Qu}}_i u_{iL}+ \bar{\tilde{D}}_{R} {Y_{Qd}}_i
  d_{iL} \right)\nn\\
 &+\Phi^\dagger \sum_{i=1}^3 \left( \bar{\tilde{U}}_{L} {Y_{Uu}}_i u_{iR}+ \bar{\tilde{D}}_{L} {Y_{Dd}}_i d_{iR} \right) + {\rm h.c.}
\end{align}
The SU(2)$_L$ symmetry relates the Yukawa couplings of LH up-type quarks to
those of the LH down-type quarks:
\begin{align}
Y_{Qu_i} = \sum_{i=1}^3 V_{u_i d_j}^* Y_{Qd_j}, \label{eq:Yukawa-SU(2)}
\end{align}
where $i=1,2,3$ and $V_{u_i d_j}$ is an element of the Cabibbo-Kobayashi-Maskawa (CKM)
matrix.

At energy scales well below the heavy vector-like quark masses, the above Yukawa couplings 
generate an effective Lagrangian for FCNC $Z'$ couplings to SM quarks, 
\begin{align}
\Delta\mathcal{L}_{\rm eff} &= -Z'_\alpha \sum_{i,j=1}^3\Big(g^L_{u_i u_j}\bar{u}_{iL}\gamma^\alpha u_{jL}+g^R_{u_i u_j}\bar{u}_{iR}\gamma^\alpha u_{jR}\nn\\
&\quad +g^L_{d_i d_j}\bar{d}_{iL}\gamma^\alpha d_{jL}+g^R_{d_i d_j}\bar{d}_{iR}\gamma^\alpha d_{jR}\Big), 
\end{align}
with
\begin{align}
 &g^L_{u_i u_j}= g'\frac{Y^*_{Qu_i}Y_{Qu_j}v_{\Phi}^2}{2m_Q^2}, ~g^R_{u_i u_j}=-g'\frac{Y^*_{Uu_i}Y_{Uu_j}v_{\Phi}^2}{2m_U^2},\nn\\
 &g^L_{d_i d_j}= g'\frac{Y^*_{Qd_i}Y_{Qd_j}v_{\Phi}^2}{2m_Q^2}, ~g^R_{d_i d_j}=-g'\frac{Y^*_{Dd_i}Y_{Dd_j}v_{\Phi}^2}{2m_D^2}. \label{effcoup}
\end{align}
Among these, the $bsZ'$ couplings $g_{sb}^L$ and $g_{sb}^R$ affect the $b\to s\mu^+\mu^-$
transitions. In particular, $g_{sb}^L$ gives a new contribution to the Wilson coefficient
of the operator $(\bar s_L\gamma^\alpha b_L)(\bar\mu \gamma_\alpha \mu)$, given by
\begin{align}
 \Delta C^\mu_9=\frac{g_{sb}^L~g'}{m_{Z'}^2}, \label{eq:C9}
\end{align}
which can explain both $P_5'$ and $R_K$ anomalies. 
If the LH $bsZ'$ coupling $g_{sb}^L$ exists, the SU(2)$_L$ relation in 
Eq.~\eqref{eq:Yukawa-SU(2)} would imply the existence of the LH $tcZ'$ coupling $g_{ct}^L$.
Unfortunately, the strength of $g_{ct}^L$ favored by the $P_5'$ and $R_K$ anomalies 
turns out to be below the discovery reach at HL-LHC, as we discuss in Sec.~\ref{senlh}.

\begin{figure}[t!]
\centering
 \includegraphics[width=.47 \textwidth]{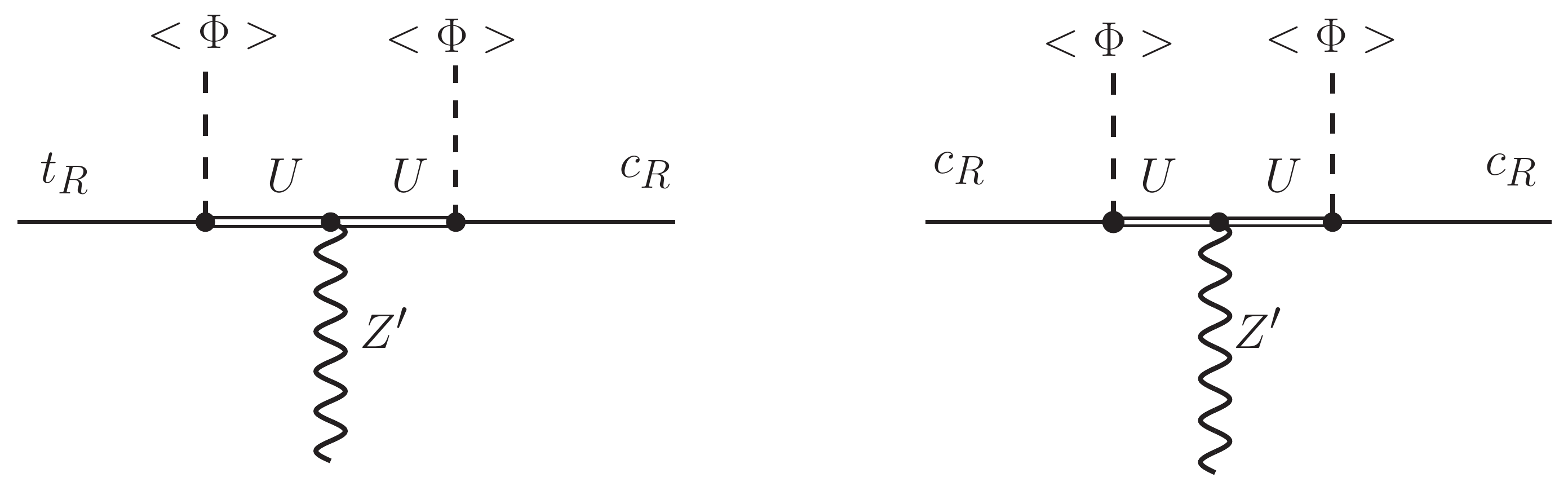}
\caption{
Feynman diagrams that generate the effective RH $tcZ'$ [left] and  $ccZ'$ [right] couplings.
}\label{vevtcz}
\end{figure}

The model, however, predicts the existence of the RH $tcZ'$ coupling $g_{ct}^R$.
The coupling is not directly linked to $B\to K^{(*)}$ transitions 
and is therefore hardly probed by $B$ and $K$ physics.
But this coupling and its effect on top physics should be viewed as on the same footing
as the  $P_5'$ and $R_K$ anomalies.
Because there is no gauge anomaly, it could even happen that the $Q$ and $D$ quarks 
are absent, or equivalently rather heavy,
but the $U$ quark could cause effects in the top/charm sector that are analogous
to the current $P_5'$ and $R_K$ ``anomalies'' in $B$ decay,
even if the latter ``anomalies'' disappear with more data.
We therefore focus on the LHC phenomenology of the RH $tcZ'$ coupling.

The RH $tcZ'$ coupling is generated by the diagram shown in the 
left panel of Fig.~\ref{vevtcz} and is given by
\begin{align}
g^R_{ct}=\left( g^R_{tc}\right)^* =-g'\frac{Y^*_{Uc}Y_{Ut}v_{\Phi}^2}{2m_U^2},
\label{eq:gctR}
\end{align}
which is nonzero only if $Y_{Uc}\neq 0$.
One sees then that the diagram in the right panel of Fig.~\ref{vevtcz} generates 
RH $ccZ'$ coupling with
\begin{align}
g^R_{cc}=-g'\frac{|Y_{Uc}|^2v_{\Phi}^2}{2m_U^2}. \label{eq:gccR}
\end{align}
This means that if the RH $tcZ'$ coupling exists, the RH $ccZ'$ coupling should also exist.
We shall therefore also consider the RH $ccZ'$ coupling for LHC phenomenology.

In short, we consider the following effective $Z'$ couplings in the collider study:
\begin{align}
\Delta\mathcal{L}_{\rm eff} 
&\supset  -g^R_{cc} \bar{c}_{R}\gamma^\alpha c_{R} Z'_\alpha
-\left( g^R_{ct} \bar{c}_{R}\gamma^\alpha t_{R} Z'_\alpha +{\rm h.c.}\right), \label{hadron}
\end{align}
with the model-dependent expressions of $g_{ct}^R$ and $g_{cc}^R$ in 
Eq.~\eqref{eq:gctR} and \eqref{eq:gccR}.
But, our collider results can be straightforwardly applied to the LH counterparts,
$g_{ct}^L$ and $g_{cc}^L$.

In principle, the model could also give the effective couplings containing the up quark, i.e.
the RH $uuZ'$, $cuZ'$ and $tuZ'$ couplings, if $Y_{Uu}$ is nonzero.
In this case, $g_{uc}^R \propto |Y_{Uu}^*Y_{Uc}|$ is constrained by $D$-meson mixing
and decays. We assume $Y_{Uu}=0$ for simplicity,
while RH $ttZ'$ coupling is discussed in Sec.~\ref{summary}.
The presence of the $U$ quark with nonzero $Y_{Ut}$ and $Y_{Uc}$ also leads to
couplings of neutral SM bosons to the $t\to c$ currents.
$tcZ$ and $tch$ couplings are induced at tree level, while $tc\gamma$ and $tc g$
couplings, forbidden at tree level due to gauge symmetry, are generated at one-loop level.
In Ref.~\cite{Altmannshofer:2014cfa}, it is claimed that the branching ratios of 
rare top quark decays induced by these FCNC couplings with the SM bosons 
are suppressed over ${\cal B}(t\to c Z')$ by roughly a loop factor,
with the latter assumed to be kinematically allowed.

We shall consider the mass range of 150 GeV $\leq m_{Z'} \leq$ 700 GeV,
where the branching ratios and total width for $Z'$ decay are nicely approximated by
\begin{align}
&\mathcal{B}(Z'\to\mu^+\mu^-) \simeq \mathcal{B}(Z'\to\tau^+\tau^-) \simeq 
 \mathcal{B}(Z'\to\nu\bar\nu) \simeq \frac{1}{3}, \notag\\
&\Gamma_{Z'} \simeq \frac{m_{Z'}^3}{4\pi v_\Phi^2} \notag\\
 &\quad~~ \simeq 0.75~{\rm GeV} \left( \frac{m_{Z'}}{150~{\rm GeV}} \right)^3
 \left( \frac{600~{\rm GeV}}{v_\Phi}\right)^2.
\end{align}
In this mass range, dominant constraints on the ($m_{Z'}, g'$) plane
comes from neutrino trident production and $B_s$ mixing~\cite{Altmannshofer:2014cfa}.
These can be recast into constraints on the VEV of the $\Phi$ field
$v_\Phi (=m_{Z'}/g')$, which can be summarized as \cite{Fuyuto:2015gmk}
\begin{align}
 0.54~{\rm TeV} \lesssim v_\Phi \lesssim 5.6~{\rm TeV}~\bigg(\frac{(34~{\rm TeV})^{-2}}{\left|\Delta C^\mu_9\right|}\bigg), \label{eq:vphi-range}
\end{align}
regardless of the value of $m_{Z'}$. 
The lower limit comes from neutrino trident production \cite{Altmannshofer:2014pba}
with $2 \sigma$ range of the CCFR result~\cite{Mishra:1991bv},\footnote{
The presence of the $Z'$ boson affects couplings of the $Z$ boson to 
the muon, tau and corresponding neutrinos via loop effect, which are constrained by 
experimental data taken at the $Z$ resonance.
A study in Ref.~\cite{Altmannshofer:2014cfa} shows that combining results
from the LEP and SLC~\cite{ALEPH:2005ab} can provide competitive or slightly better 
limits than the CCFR for $m_{Z'}\gtrsim600$ GeV.
}
while the upper limit is set by $B_s$ mixing~\cite{Crivellin:2015mga}
with BSM effects allowed within $15\%$ and the assumption of $m_Q \lesssim 10$ TeV.
The upper limit becomes tighter for larger $m_Q$, 
e.g. $v_\Phi \lesssim 5.4~(3.9)~{\rm TeV} \times [(34~{\rm TeV})^{-2}/|\Delta C_9^\mu|]$ 
for $m_Q=20~(50)$ TeV.

It is convenient to introduce the mixing parameters~\cite{Fuyuto:2015gmk} between
vector-like quark $U$ and RH top or charm quark defined by
\begin{align}
 \delta_{Uq} \equiv \frac{Y_{Uq} v_\Phi}{\sqrt{2} m_U}, \quad (q=t, c). \label{eq:deltaUq}
\end{align}
Small mixing parameters are assumed in obtaining the effective couplings of Eq.~\eqref{eq:gctR}
and \eqref{eq:gccR}.
In the following analysis, we allow the mixing strengths up to 
the Cabibbo angle, i.e. $|\delta_{Ut}|, |\delta_{Uc}| \leq \lambda \simeq 0.23$,
and the RH $tcZ'$ coupling is constrained as
\begin{align}
|g_{ct}^R| &= \frac{m_{Z'}}{v_\Phi}|\delta_{Uc}||\delta_{Ut}|  \notag\\
&\lesssim 0.013\times \left( \frac{m_{Z'}}{150~{\rm GeV}} \right) 
 \left( \frac{600~{\rm GeV}}{v_\Phi}\right). \label{eq:gctR-range}
\end{align}
If the Yukawa couplings are hierarchical, e.g., $|Y_{Ut}| \gg |Y_{Uc}|$, this is further suppressed
by $|Y_{Uc}/Y_{Ut}|$.
A similar constraint holds for $g_{cc}^R$.
These set the target ranges for the LHC study.

%
%

\section{Search for $Z'$ at the LHC}
\label{direct}

\subsection{ $tZ'$ and $\bar{t}Z'$ processes }
\label{zptc}

\begin{figure}[t!]
\centering
\includegraphics[width=.45 \textwidth]{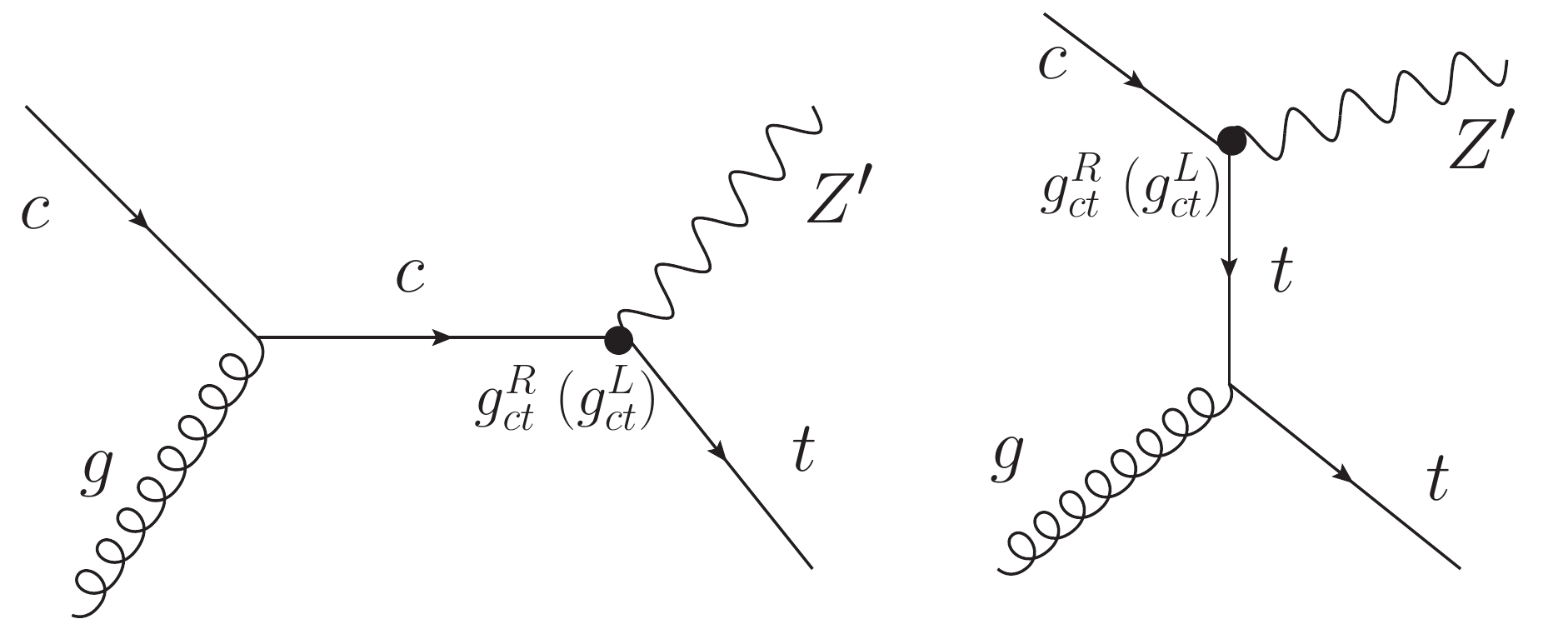}
\caption{Feynman diagrams contributing to $p p \to t Z'$.}
\label{tcz}
\end{figure}
The RH $tcZ'$ coupling in Eq.~\eqref{hadron} generates the parton-level process 
$cg\to tZ'$ through the Feynman diagrams in Fig.~\ref{tcz}, leading to $pp\to tZ'$ at the LHC.
We assume subsequent decays of $Z'\to \mu^+\mu^-$ and
$t\to bW^+ (\to \nu_\ell \ell^+)$ with $\ell =e$ or $\mu$.
In this subsection, we study the FCNC-induced process 
$pp \to tZ' \to b \nu_\ell \ell^+ \mu^+ \mu^-$ ($tZ'$ process) and its conjugate process 
$pp \to \bar{t}Z' \to \bar{b}\bar{\nu}_\ell\ell^- \mu^+ \mu^-$ ($\bar tZ'$ process) at the 
14 TeV LHC, and analyze the prospect of discovering such a $Z'$ boson.
The RH $tcZ'$ coupling also generates processes with an extra charm quark in
the final states, i.e. $gg \to t\bar cZ'$ or $\bar tcZ'$.
We will veto extra jets in the following analysis, but the latter processes can contribute
to the signal region if the charm jet escapes detection.
Hence, we also include the contributions from $gg \to t\bar cZ'/\bar tcZ'$ as a signal.


For sake of our collider analysis, we take two benchmark points
for the effective theory defined by Eq.~\eqref{hadron}:
\begin{itemize}
\item Case A:  $\left|g^R_{ct}\right|=0.01$, $m_{Z'}=150$ GeV;
\item Case B:   $\left|g^R_{ct}\right|=0.01$, $m_{Z'}=200$ GeV.
\end{itemize}
In Case A, where $m_{Z'}<m_t$, the $t\to cZ'$ decay is kinematically allowed
with $\mathcal{B}(t\to cZ')\simeq 2\times 10^{-5}$,
and it contributes to $gg\to t \bar cZ'/\bar tcZ'$ via $gg\to t \bar t$.
On the other hand, in Case B with $m_{Z'}> m_t$, the $t\to cZ'$ decay is 
kinematically forbidden.\footnote{
For $m_{Z'} > m_{t}$, a three-body decay $t\to c \mu^+\mu^-$ may still happen
through an off-shell $Z'$, and can contribute to the signal region via the $t\bar t$ events.
In this case, the $Z'$ mass cannot be reconstructed from the dimuon invariant mass, 
but the top quark mass reconstruction may help discriminate signal and backgrounds.
In Case B with $m_{Z'}=200$ GeV, such a contribution is very tiny and is not included
in our analysis, although it could be important for a $Z'$ mass nearby the top quark mass.
}
Moreover, behavior of event distributions for SM backgrounds is qualitatively different 
depending on whether the $Z'$ mass is below or above the top-quark mass.
The coupling value is in the range of Eq. (\ref{eq:gctR-range})
implied by the gauged $L_\mu -L_\tau$ model.

The signal cross sections are proportional to $|g_{ct}^R|^2\times
\mathcal{B}(Z'\to \mu^+\mu^-)$ if the $Z'$ width is narrow. 
We assume $\mathcal{B}(Z'\to \mu^+\mu^-)= 1/3$, motivated by the gauged $L_\mu -L_\tau$ model, 
and $\Gamma_{Z'}\lesssim 1$ GeV for each case.
Besides these assumptions, the analysis in this subsection is model-independent.
Effects of different $Z'$ branching ratios can be taken into account by
rescaling $|g_{ct}^R|$.

A similar BSM process $pp\to tZ \to \ell\nu b \ell^+\ell^-$ induced by $tcZ$ couplings 
has been studied by the CMS experiment with 8 TeV data~\cite{Sirunyan:2017kkr}.
Our study closely follows this analysis. There exist several non-negligible SM 
backgrounds for the signal $b \nu_\ell \ell^+ \mu^+ \mu^-$ ($tZ'$ process) 
and $\bar{b}\bar{\nu}_\ell\ell^- \mu^+ \mu^-$ ($\bar tZ'$ process):
\begin{itemize}
\item \textbf{$tZj$ and $\bar{t}Zj$ backgrounds}: 
 The $tZj$ background predominantly originates from
\begin{align}
\qquad u + b \to t + Z + d~~\mbox{or}~~\bar{d} + b \to t + Z + \bar{u},
\end{align}
 with smaller contributions from $c$- or $\bar{s}$-initiated processes,
 while $\bar{t}Zj$ is generated by the charge-conjugate processes
\begin{align}
\qquad d + \bar{b} \to \bar{t} + Z + u~~\mbox{or}~~\bar{u} + \bar{b} \to \bar{t} + Z + \bar{d}.
\end{align}
 The $tZj$ cross section is larger than $\bar{t}Zj$, as the parton distribution 
 function (PDF) of the $u$ quark is larger than the $d$ quark in $pp$ 
 collisions~\cite{Campbell:2013yla}. 
 Thus, the $tZ'$ process suffers from larger background. 
\item \textbf{$t\bar{t}Z$ background}: 
 $t\bar{t}Z$ becomes background for the $\bar{t}Z'$ ($tZ'$) process, 
 if the $t$ ($\bar{t}$) decays hadronically and the $\bar{t}$ ($t$) decays leptonically, i.e.
 $t\bar{t}Z \to (bq\bar{q}')(\bar{b} \bar{\nu}_\ell \ell^-)(\mu^+\mu^-)$
 [$(\bar{b}q'\bar{q})(b \nu_\ell \ell^+)(\mu^+\mu^-)$],
 with some of the jets undetected.
 Indeed, $t\bar{t}Z$ constitutes a major part of the overall background.
%
\item \textbf{$t\bar{t}W$ background}: 
 $t\bar{t}W$ is another leading source of background. If the $t$, $\bar{t}$ and $W$ 
 all decay leptonically and a jet goes undetected, it can give the event topology with 
 trilepton ($\mu^+\mu^-\ell$), missing transverse energy ($\cancel{E}_{T}$) and 
 $b$-tagged jet. 
The $t\bar{t}W^+$ production cross section  is larger than 
$t\bar{t}W^-$~\cite{Campbell:2012dh} for $pp$ collisions.
 Thus, the $tZ'$ process again suffers larger background. 
%
\item \textbf{$WZ$+heavy-flavor jets and $WZ$+light jets}:
 The $WZ$ or $W\gamma^*$ production in association with heavy-flavor (h.f.) or light jets 
 also contribute to background, if both $W$ and $Z/\gamma^*$ decay leptonically and 
 a jet gets misidentified as a $b$-tagged jet. 
 Here, the h.f. jet refers to the $c$-jet. The rejection factors for the $c$-jet and the light jet 
 are taken to be $5$ and $130$, respectively~\cite{ATLAS:2014ffa}. 
 The cross section for $W^+Z$+light jets is larger than $W^-Z$+light jets, 
 while the $W^\pm Z$ production cross sections in association with h.f.-jets are identical. 
 This also gives larger background to the $tZ'$ process than $\bar t Z'$.
\end{itemize}

We do not consider processes such as $t\bar{t}$, Drell-Yan (DY), $W$+jets, 
which could contribute to background if one or two nonprompt leptons are
produced and reconstructed. 
These backgrounds are not properly modeled in simulation and require data for 
better estimation. The analysis of the similar process $pp\to tZ$ by CMS~\cite{Sirunyan:2017kkr} 
shows that such processes provide subdominant contributions to the total background.
In the case of the $tZ'$ process, stricter cuts on the transverse momenta of the muons 
may reduce such contributions. These are beyond the scope of this paper.

\begin{figure}[t!]
\centering
\includegraphics[width=.49 \textwidth]{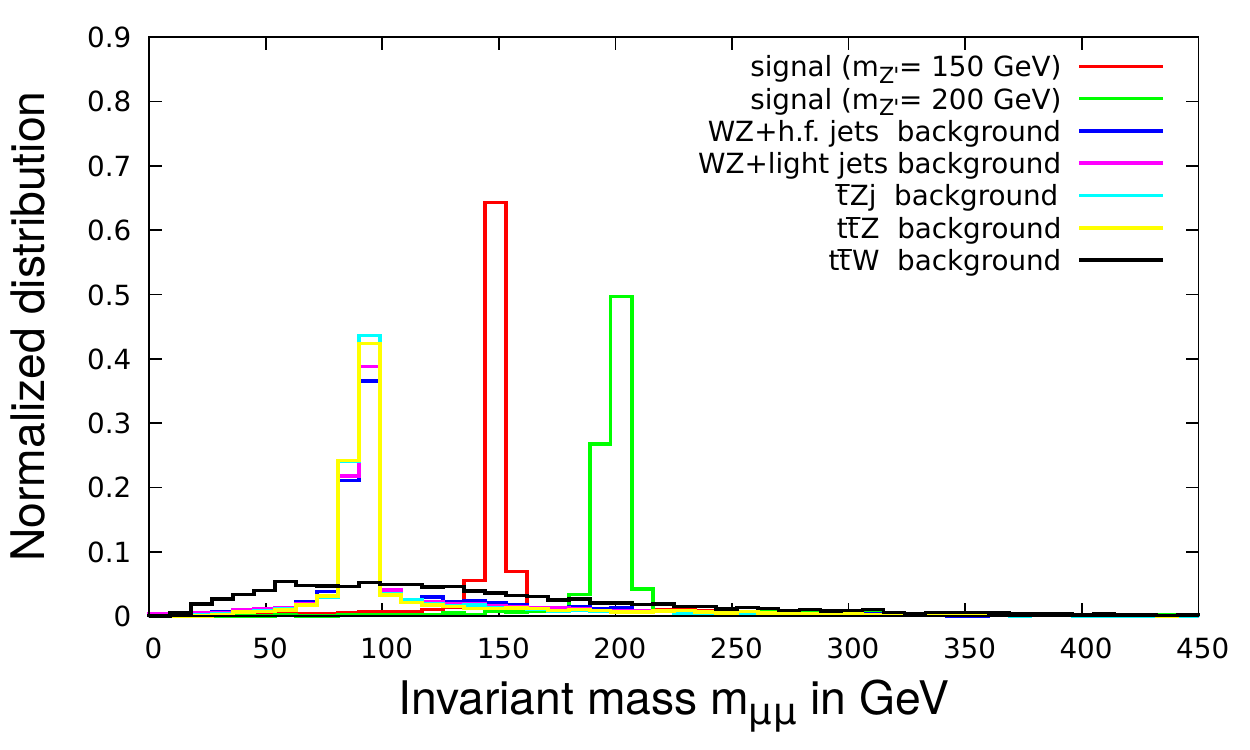}
\caption{Normalized distributions of the dimuon invariant mass for the $\bar tZ'$ process 
in Case A ($m_{Z'}=150$ GeV) and B ($m_{Z'}=200$ GeV), and for the corresponding backgrounds,
with close-to-default cuts in MadGraph.
}
\label{mass_zpt}
\end{figure}

The signal and background samples are generated at leading order (LO) in the $pp$ collision with center of mass energy $\sqrt{s}=14$ TeV, by the Monte Carlo event generator MadGraph5\_aMC@NLO~\cite{Alwall:2014hca}, interfaced to PYTHIA~6.4~\cite{Sjostrand:2003wg} for showering. 
To include inclusive contributions, we generate the matrix elements of signal and backgrounds with up to one additional jet in the final state, followed by matrix element 
and parton shower merging with the MLM matching scheme~\cite{Alwall:2007fs}. 
Due to computational limitation, we do not include processes with two or more additional 
jets in the final state.
The event samples are finally fed into the fast detector simulator
Delphes~3.3.3~\cite{deFavereau:2013fsa} for inclusion of (ATLAS-based) detector effects. 
The effective theory defined by Eqs.~\eqref{lepton} and \eqref{hadron} is implemented by
FeynRules~2.0~\cite{Alloul:2013bka}. We adopt the PDF set CTEQ6L1~\cite{Pumplin:2002vw}.
The LO $\bar{t}Zj$ and $t\bar{t} Z$ cross sections are normalized to the next-to-leading
order (NLO) ones by $K$-factors of 1.7 and 1.56, respectively~\cite{Campbell:2013yla}. 
For simplicity, we assume $tZj$ has the same NLO $K$-factor as $\bar{t}Zj$.
The NLO $K$-factor for the $t\bar{t} W^-$ ($t\bar{t} W^+$) process is taken to be
1.35 (1.27)~\cite{Campbell:2012dh}. The LO cross section for the $W^-Z$+light jets 
background is normalized to the next-to-next-to-leading order (NNLO) one by a factor 
of 2.07~\cite{Grazzini:2016swo}. We assume the same correction factor for 
$W^+Z$+light jets and $W^\pm Z$+h.f. jets for simplicity.

The signal cross sections for the $tZ'$ and $\bar{t}Z'$ processes are identical, while some 
of the dominant (and the total) background cross sections are smaller for the latter process.
The $\bar{t}Z'$ process is, therefore, better suited for discovering the $Z'$.
It turns out that combining the $tZ'$ and $\bar{t}Z'$ processes can improve 
discovery potential.
In the following, we primarily investigate the $\bar tZ'$ process in showing details of 
our analysis, and finally give combined results of the $tZ'$ and $\bar{t}Z'$ processes.

\begin{figure*}[htb!]
\centering
\includegraphics[width=.49 \textwidth]{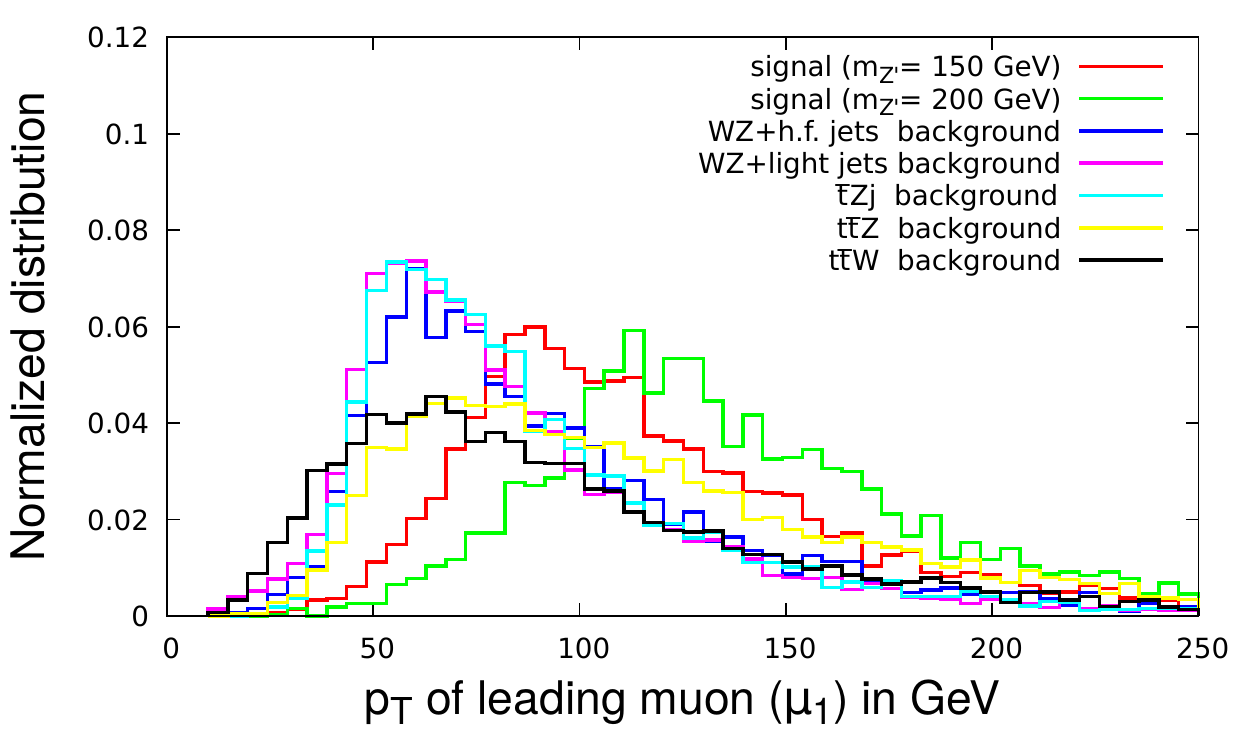}
\includegraphics[width=.49 \textwidth]{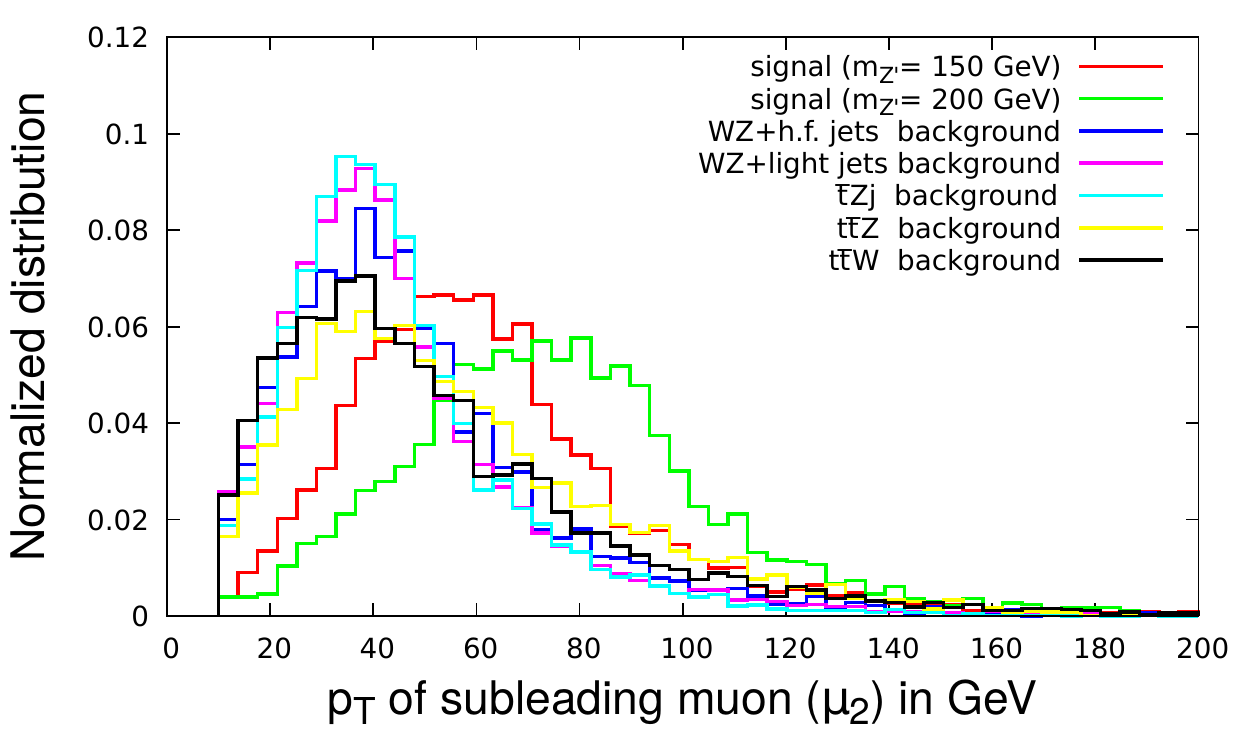}
\caption{Normalized $p_T$ distributions
for the leading [left] and subleading [right] muons, 
for the $\bar{t}Z'$ process and  its backgrounds as in Fig.~\ref{mass_zpt}.}
\label{ptlep_zpt}
\end{figure*}

\begin{figure*}[htb!]
\centering
\includegraphics[width=.49 \textwidth]{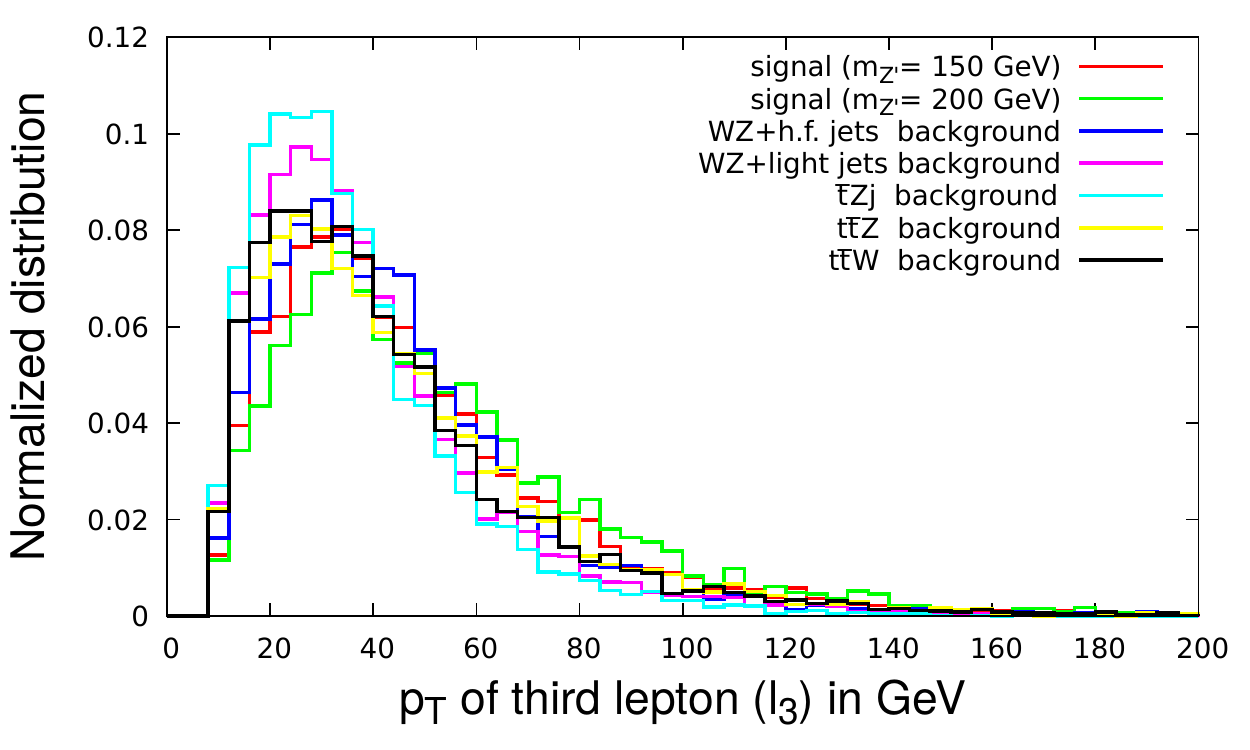}
\includegraphics[width=.49 \textwidth]{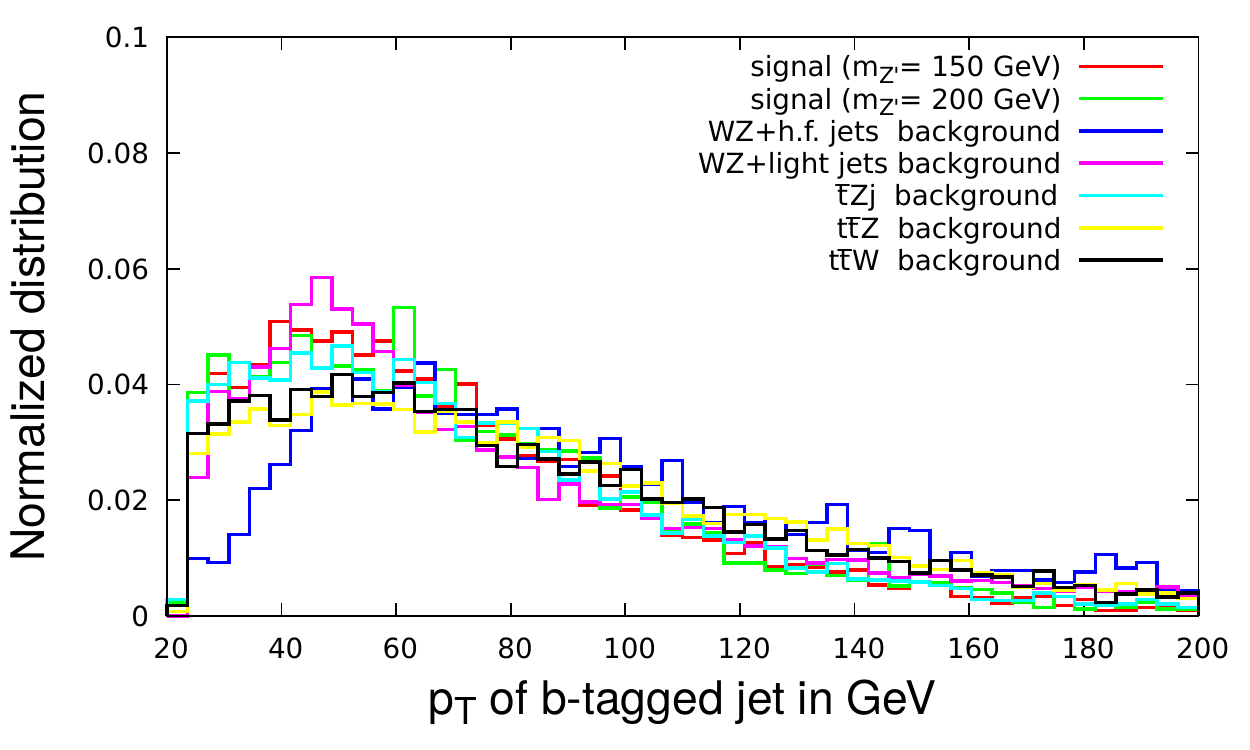}
\caption{Normalized $p_T$ distributions for the third lepton [left] and 
$b$-tagged jet [right], for the $\bar{t}Z'$ process and its backgrounds 
as in Fig.~\ref{mass_zpt}.}
\label{ptbl_zpt}
\end{figure*}

We present, in Fig.~\ref{mass_zpt}, the normalized event distributions of the 
dimuon invariant mass $m_{\mu\mu}$ for the $\bar t Z'$ process in Case A and B, 
and for the corresponding background contributions.
The distributions are obtained by applying default cuts in MadGraph 
with minor modifications. 
In Figs.~\ref{ptlep_zpt} and \ref{ptbl_zpt}, the normalized $p_T$ distributions are similarly
shown for the leading and subleading muons, the third lepton and the $b$-tagged jet, respectively.


\begin{table*}
\centering
\begin{tabular}{|c|c|c|c|c|c|c|c|c|}
\hline
\hline
Cuts               & Signal (Case A)        &$\bar{t}Z j$     & $t\bar{t} Z$  &$t\bar{t} W^-$    & $W^-Z$+light jets   & $W^-Z$+h.f. jets & Total BG \\ 
\hline
\hline
\textbf{Pre-selection cuts}   & 0.410           & 0.872 (1.552)   & 1.672         & 0.514 (1.384)  & 0.641 (0.868)       &  4.55   & 8.25 (10.03) \\ 
&&&&&&&\\
\textbf{Selection cuts}      & 0.090             & 0.012 (0.022)   & 0.026         & 0.023 (0.071)  & 0.012 (0.015)     &   0.017  & 0.090 (0.151)  \\
\textbf{(No jet veto)}&&&&&&&\\
&&&&&&&\\
\textbf{Selection cuts}      & 0.085             & 0.011 (0.020)   & 0.014         & 0.014 (0.039)  & 0.005 (0.007)      &  0.014  & 0.058 (0.094)   \\

\hline
\end{tabular}
\caption{Effects of two sets of cuts on cross sections (in fb) for 
the $\bar{t} Z'$ and SM background processes in Case A ($m_{Z'}=150$ GeV).
The effect of the selection cuts without the subleading jet veto is also shown.
(See text for details.)
The second column gives the signal process, 
while the effects on individual backgrounds are tabulated in third to seventh columns. 
Cross sections for backgrounds of the conjugate process $t Z'$ are given
in parentheses, if they differ from the case of the $\bar t Z'$ process:
$tZj$ (third column), $t\bar{t} W^+$ (fifth column) and $W^+Z$+light jets (sixth column),
where similar sets of cuts as the $\bar t Z'$ process are applied.
The last column shows the sum of all background cross sections.}
\label{cut_table_zpt150}
\end{table*}
\begin{table*}
\centering
\begin{tabular}{|c|c|c|c|c|c|c|c|c|}
\hline
\hline
Cuts                        & Signal (Case B)       &$\bar{t}Z j$     & $t\bar{t} Z$  &$t\bar{t} W^-$    & $W^-Z$+light jets   & $W^-Z$+h.f. jets  & Total BG \\
\hline
\hline
\textbf{Pre-selection cuts} &  0.186       & 0.872 (1.552)    & 1.672          & 0.514 (1.384)     & 0.641 (0.868)      &  4.55  & 8.25 (10.03)  \\
&&&&&&&\\
\textbf{Selection cuts}     &  0.040       & 0.006 (0.010)  & 0.014          & 0.012 (0.035)     & 0.005 (0.007)        &  0.008 & 0.045 (0.074) \\
\textbf{(No jet veto)}&&&&&&&\\
&&&&&&&\\
\textbf{Selection cuts}     &  0.037       & 0.005 (0.009)    & 0.007          & 0.008 (0.021)     & 0.002 (0.003)      &  0.007 & 0.029 (0.047) \\
\hline
\end{tabular}
\caption{Same as Table \ref{cut_table_zpt150}, but for Case B ($m_{Z'}=200$ GeV).
}\label{cut_table_zpt200}
\end{table*}

We use two sets of cuts on the signal and background processes as explained below.

\noindent\textbf{Pre-selection cuts}: 
This set of cuts is used at the generator level. The leading, subleading and third leptons 
in an event are required to have minimum $p_T$ of 60 GeV, 30 GeV and 15 GeV, 
respectively, in both Case A and B. The maximum pseudo-rapidity of all leptons 
are required to be $ |\eta^\ell| < 2.5$. The transverse momentum of jets are required 
to be greater than 20 GeV. The minimum separation between the two oppositely-charged
muons are required to be $\Delta R > 0.4$. 
The rest of the cuts are set to their default values in MadGraph.

\noindent\textbf{Selection cuts}:
Utilizing the signal and background distributions in Figs.~\ref{mass_zpt}, 
\ref{ptlep_zpt} and \ref{ptbl_zpt}, we impose a  further set of cuts.
Events are selected such that each should contain three (at least two muon type)
leptons and at least one $b$-tagged jet. Jets are reconstructed by the anti-$k_T$
algorithm with radius parameter $ R = 0.5$. Stricter cuts on 
lepton transverse momenta are applied: the leading muon, subleading muon and third lepton 
in an event are required to have minimum $p_T$ of 60 (75) GeV, 30 (45) GeV 
and 20 (20) GeV, respectively, in Case A (B). 
The third lepton is assumed to arise from the top-quark decay accompanied by
missing transverse energy $\cancel{E}_{T}$ and $b$ jet. 
We require that $\cancel{E}_{T} > 30$ GeV and the reconstructed $W$ boson mass 
$m^W_T >10$ GeV. The leading $b$-tagged jet is required to have $p_T > 20$ GeV.
An event is rejected if the $p_T$ of the subleading jet or subleading $b$-tagged jet is 
greater than $20$ GeV. 
This veto significantly reduces the $t\bar{t}Z$ and $t\bar{t}W$ backgrounds, as both 
processes contain two $b$-jets from decay of the $t$ and $\bar{t}$.
We will also analyze the impact of removing such a jet veto shortly.
We finally apply the invariant mass cut $|m_{\mu\mu} -m_{Z'}|<15$ GeV on two
oppositely-charged muons.
If an event contains three muons, there are two ways to make a pair of two 
oppositely-charged muons. 
In such a case, we identify the pair having the invariant mass $m_{\mu\mu}$ closer 
to $m_{Z'}$ as the one coming from the $Z'$ decay, and impose the above invariant mass
cut on this pair.
%

%


The effects of these two sets of cuts on the signal and background processes are
illustrated in Table~\ref{cut_table_zpt150} for Case A, and Table~\ref{cut_table_zpt200} 
for Case B. From these tables, we see that the selection cuts
significantly reduce the number of background events ($B$), and the number of
signal events ($S$) becomes larger than $B$ for the $\bar tZ'$ process in both 
Case A and B. The expected numbers of events with integrated luminosity 
$\mathcal{L}= 300$ fb$^{-1}$ are $S\simeq 26~(11)$ and $B\simeq 17~(9)$
in Case A (B) for the $\bar t Z'$ process.
For comparison, the effects of removing the veto on the subleading jet are also shown
in the tables.
Without the jet veto, the signal events slightly increases as $S\simeq 27~(12)$,
but the background events increase more as $B\simeq 27~(14)$  
in Case A (B) for the $\bar t Z'$ process.
This illustrates the advantage of imposing the veto on the subleading jet.

To estimate the signal significance, we use~\cite{Cowan:2010js}
\begin{align}
\mathcal{Z} = \sqrt{2 \left[ (S+B)\ln\left( 1+S/B \right)-S \right]}. \label{eq:signif}
\end{align}
This becomes the well-known $\mathcal{Z}\simeq S/\sqrt{B}$ form for $S\ll B$,
but it does not hold in the current case.
We require $\mathcal{Z}\geq 5$ for 5$\sigma$ discovery.
In Case A (B), therefore, the $Z'$ can be discovered at $5\sigma$ in the $\bar{t}Z'$ process 
with integrated luminosity $\mathcal{L}=290~(730)$ fb$^{-1}$.
Discovery in the $tZ'$ process would require more data:
$\mathcal{L}=410~(1060)$ fb$^{-1}$ in Case A (B).
Combining the $tZ'$ and $\bar tZ'$ processes, one could discover the $Z'$ 
with lower integrated luminosities: $\mathcal{L}=180~(450)$ fb$^{-1}$ in Case A (B).
Therefore, better discovery potential is attained with the combined $tZ'$ and 
$\bar{t} Z'$ processes.
In the following, we will give results for this combined case and also call it
$tZ'$ process collectively if there is no confusion.

Before closing this subsection, we briefly discuss the use of some existing LHC data to search for the $tcZ'$ coupling.
CMS~\cite{Sirunyan:2017kkr} has studied the SM process $pp\to tZq$ in the three lepton (electron or muon) final state with 8 TeV data, measuring the cross section of $\sigma(pp\to tZq \to \ell \nu b\ell^+\ell^- q) = 10^{+8}_{-7}$ fb, which is consistent with SM prediction of 8.2 fb.
Taking this as background, CMS has also searched for the BSM process $pp\to tZ$ induced by $tqZ$ ($q=u,c$) couplings; no evidence was found, resulting in the 95\% CL upper limits of ${\cal B}(t\to uZ)<0.022$\% and ${\cal B}(t\to cZ)<0.049$\%.

In the CMS analysis~\cite{Sirunyan:2017kkr}, $tZq$ 
production has been searched for with the invariant mass 
cut of $76$ GeV $< m_{\ell\ell} < 106$ GeV on two 
oppositely-charged same-flavor leptons. Hence, the search is sensitive to the $tZ'$ process if the $Z'$ mass falls into this window.
The measured cross section for the three muon 
channel is $\sigma(pp\to tZq \to \mu \nu b\mu^+\mu^- q) = 5^{+9}_{-5}$ fb, 
while the SM prediction is around 2.1 fb with an uncertainty of less than 10\%.
Following the same event selection cuts as the CMS analysis, we calculate 
the $Z'$ contribution to be 17.4 fb$\times |g_{ct}^R/0.05|^2$ for $m_{Z'}=95$ GeV by MadGraph 
followed by showering and incorporating CMS based detector effects.
Symmetrizing the experimental uncertainties by naive average and allowing the $Z'$ effect 
to enhance the cross section up to 2$\sigma$ of the measured value, we obtain 
an upper limit of $|g^R_{ct}| \lesssim 0.05$ for $m_{Z'}\sim m_{Z}$.

\begin{figure}[b!]
\centering
\includegraphics[width=.49 \textwidth]{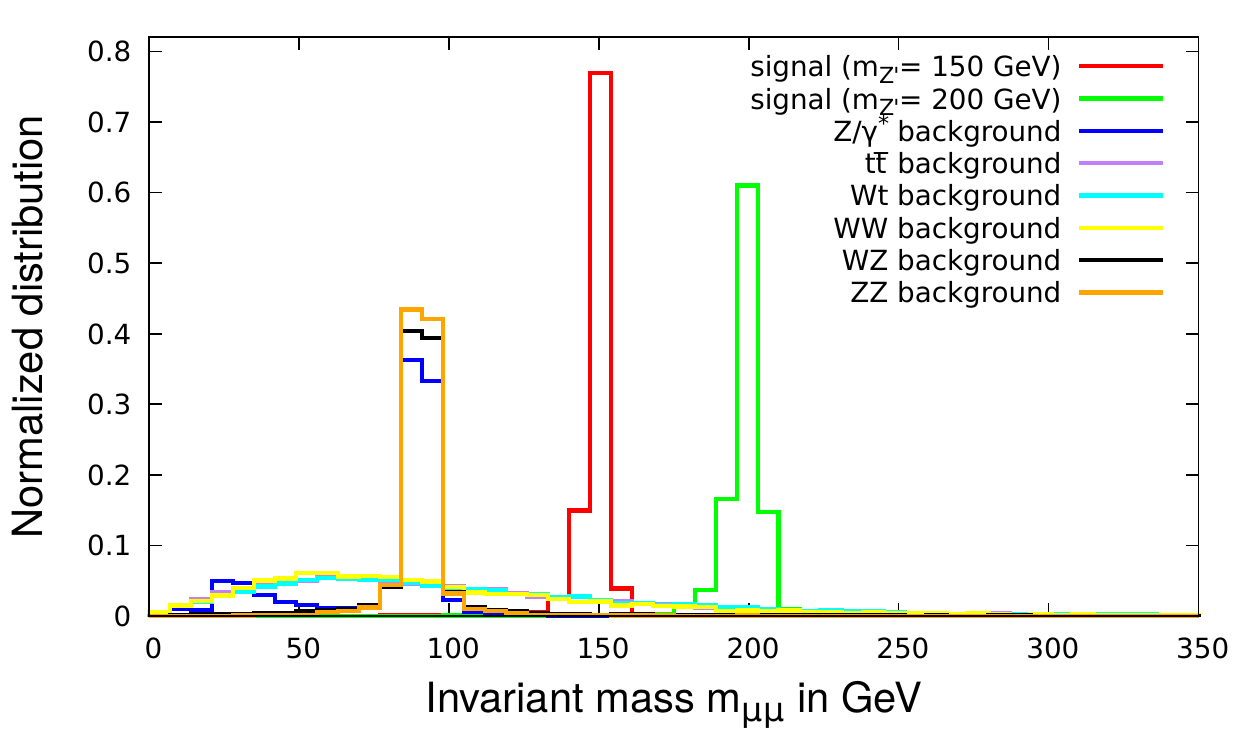}
\caption{Normalized distributions of dimuon invariant mass for the dimuon process 
$p p \to Z' + X \to \mu^+\mu^- + X$ in Case I ($m_{Z'}=150$ GeV), II ($m_{Z'}=200$ GeV)
and for the backgrounds, with close to default cuts in MadGraph.}
\label{invmass}
\end{figure}
\begin{figure*}[t!]
\centering
\includegraphics[width=.49 \textwidth]{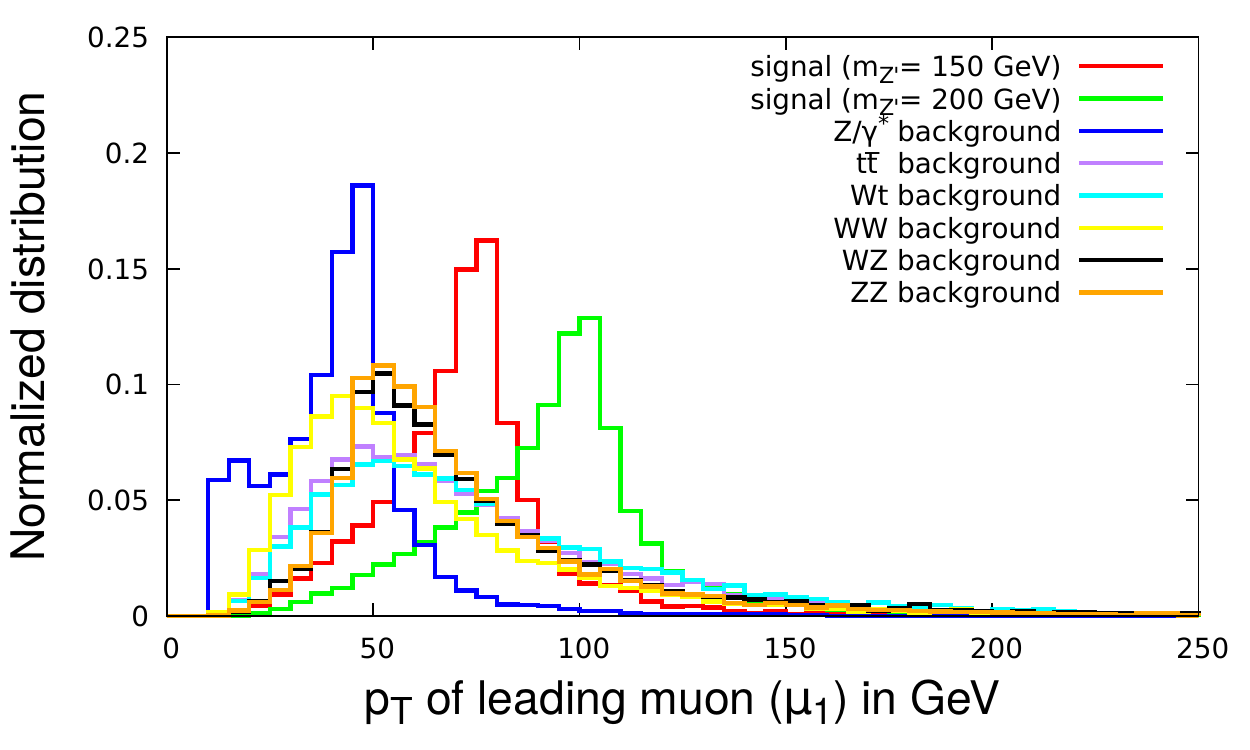}
\includegraphics[width=.49 \textwidth]{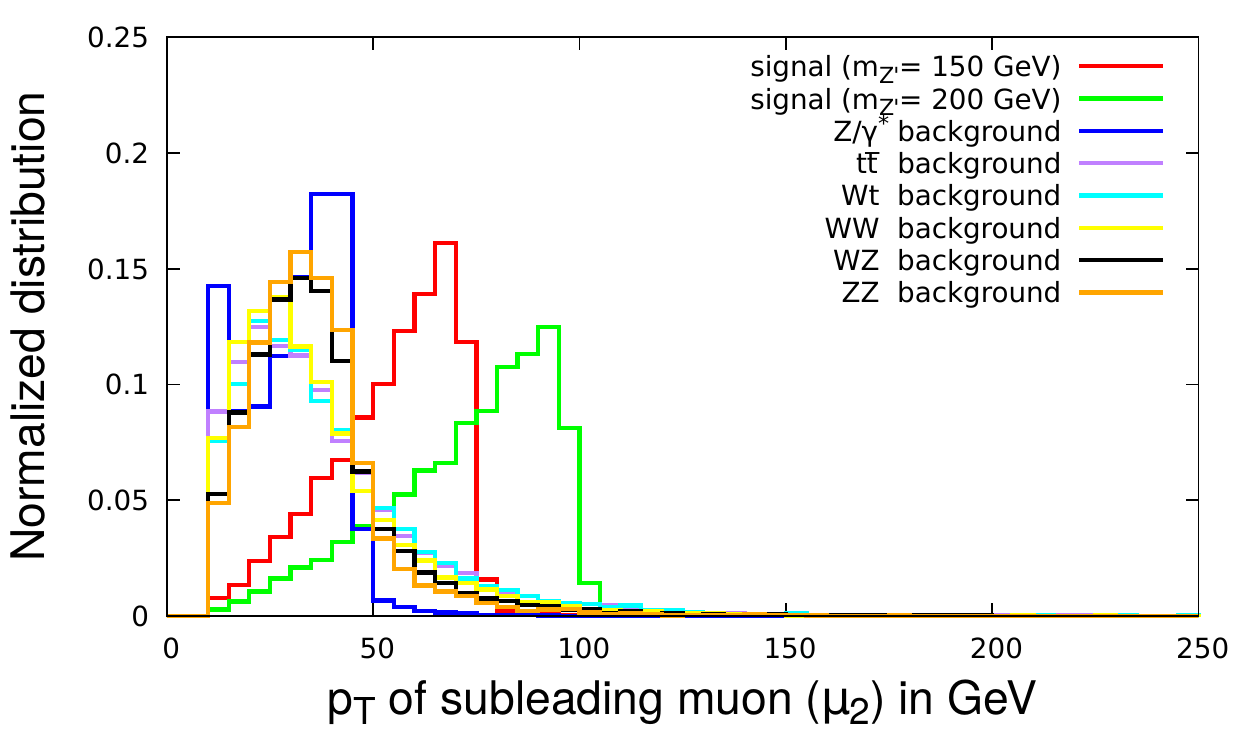}
\caption{Normalized $p_T$ distributions for leading 
[left] and subleading [right] muons, for the dimuon process and its backgrounds
as in Fig.~\ref{invmass}.}
\label{ptdist}
\end{figure*}

\subsection{Dimuon process: $pp \to Z'+ X \to \mu^+ \mu^-+ X $}
\label{dimuonc}

The flavor conserving $ccZ'$ coupling $g_{cc}^R$ in Eq.~\eqref{hadron} gives rise to 
the parton-level process $c\bar c\to Z'$. 
Thus, the $Z'$ can also be searched for 
via $pp \to Z' + X \to \mu^+ \mu^- + X$ (dimuon process), where
existing dimuon resonance search results at the LHC can already constrain $|g_{cc}^R|$.
The experimental searches do not veto extra activities $X$; hence, we also include
subdominant contributions from $cg \to cZ'$ and $gg\to c \bar{c} Z'$ processes,
induced by the RH $ccZ'$ coupling.
In the following analysis, we adopt the 13 TeV results by ATLAS~\cite{ATLAS:2016cyf}
and CMS~\cite{CMS:2016abv} (both based on $\sim$ 13 fb$^{-1}$ data). 
The ATLAS analysis puts 95\% CL upper limits on $Z'$ production cross section times $Z'\to\mu^+\mu^-$ branching ratio for 150 GeV $\lesssim m_{Z'} \lesssim$ 5 TeV, 
while the CMS analysis provides 95\% CL upper limits on the quantity $R_\sigma$, which is defined as the ratio of dimuon production cross section via $Z'$ to the one 
via $Z$ or $\gamma^*$ (in the dimuon-invariant mass window of 60--120 GeV),
for 400 GeV $\lesssim m_{Z'} \lesssim$ 4.5 TeV.
We interpret the latter as the limits on
\begin{align}
R_\sigma = \frac{ \sigma(pp\to Z'+ X)\mathcal{B}(Z'\to\mu^+\mu^-) }{ \sigma(pp\to Z + X)\mathcal{B}(Z\to\mu^+\mu^-) }, 
\end{align}
and convert them into the limits on $\sigma(pp\to Z' +X)\mathcal{B}(Z'\to\mu^+\mu^-)$ by
multiplying the SM prediction of $\sigma(pp\to Z + X)\mathcal{B}(Z\to\mu^+\mu^-)=1928.0$ 
pb \cite{CMS:2015nhc}. 
With parameters allowed by these searches, we study the prospect of discovering 
the dimuon process at the 14 TeV LHC.

\begin{figure*}[t!]
\centering
  \raisebox{1.2mm}{\includegraphics[width=.38\linewidth]{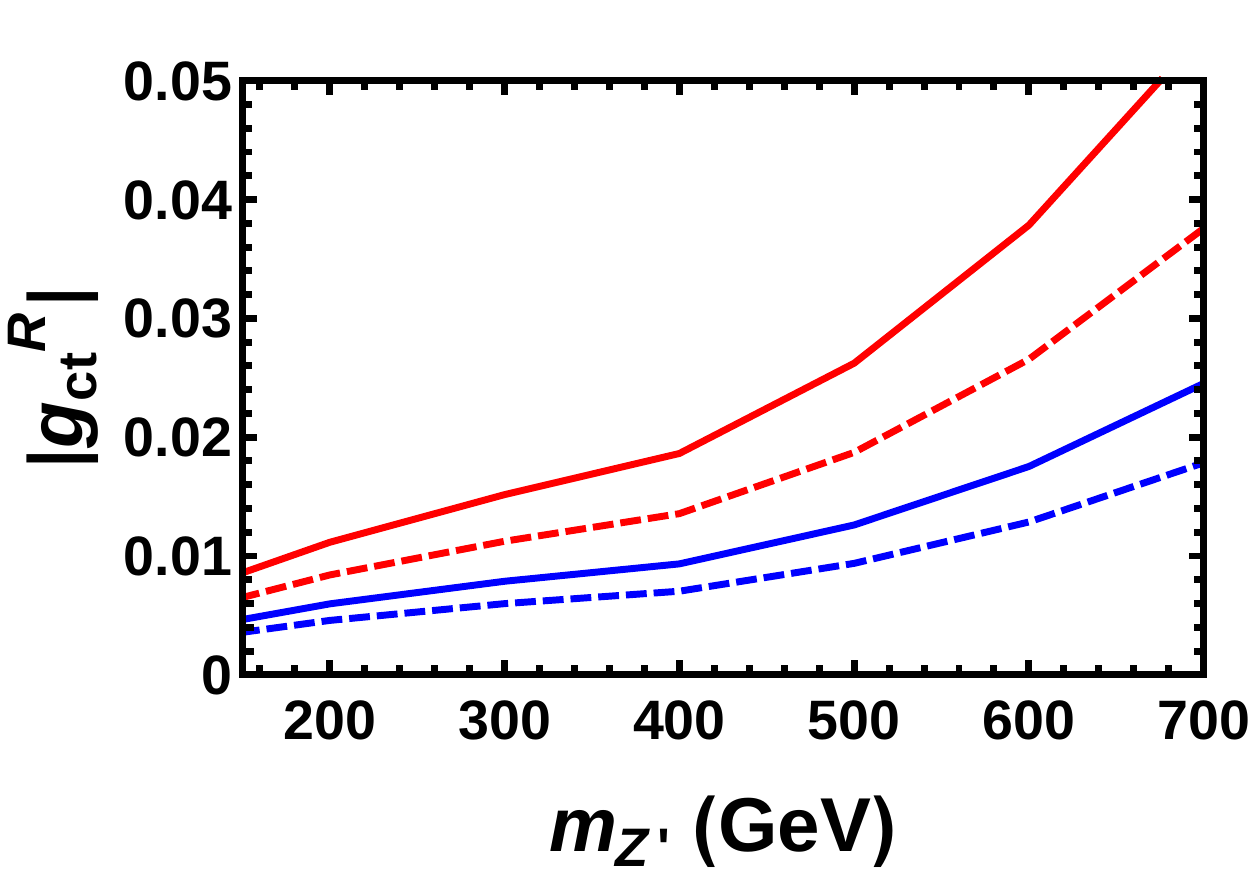}}
  \includegraphics[width=.582\linewidth]{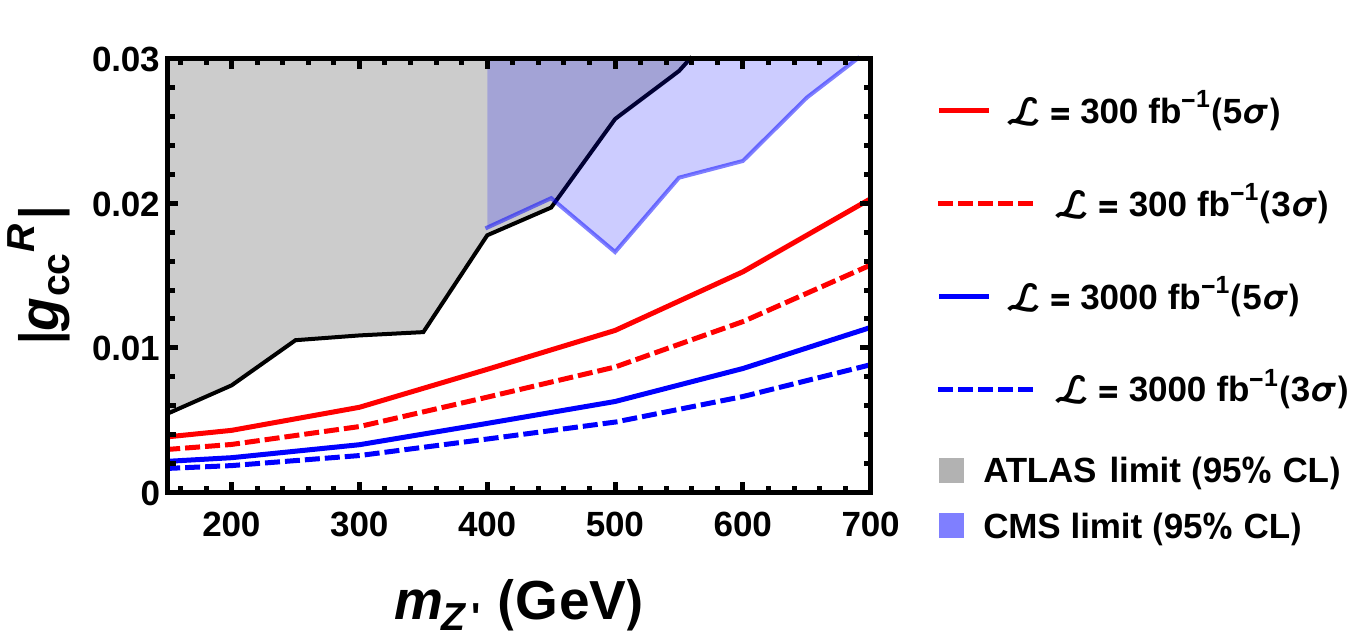}
\caption{
[Left] 5$\sigma$ discovery reach in $|g_{ct}^R|$ strength vs $m_{Z'}$ for the combination
of the $pp \to t Z'$ and $\bar t Z'$ processes at the 14 TeV LHC with 300 fb$^{-1}$ 
(upper red solid line) or 3000 fb$^{-1}$ (lower blue solid line) data, 
and corresponding 3$\sigma$ reach shown by dashed lines. 
[Right] Same as  left panel, but for $|g_{cc}^R|$ vs $m_{Z'}$  for the dimuon process 
$pp\to Z'+X \to \mu^+\mu^- +X$. 
The gray (semi-transparent blue) shaded region is excluded 
at 95\% CL by the dimuon resonance search of ATLAS \cite{ATLAS:2016cyf} 
(CMS \cite{CMS:2016abv}) with $\sim 13$ fb$^{-1}$ data at the 13 TeV LHC.
$\mathcal{B}(Z'\to\mu^+\mu^-)=1/3$ and $\Gamma_{Z'}/m_{Z'}\lesssim 1$\% are 
assumed in both panels.
}\label{eff}
\end{figure*}

As in the previous subsection, we choose two benchmark points:
\begin{itemize}
\item Case I:   $\left|g^R_{cc}\right|=0.005$, $m_{Z'}=150$ GeV;
\item Case II:  $\left|g^R_{cc}\right|=0.005$, $m_{Z'}=200$ GeV.
\end{itemize}
We assume narrow $Z'$ width ($\Gamma_{Z'}\lesssim 1$ GeV) and 
$\mathcal{B}(Z'\to \mu^+ \mu^-)= 1/3$ for each case.
The benchmark points are allowed by the dimuon resonance searches,
as can be seen from the right panel of Fig.~\ref{eff} in the next section.

For treatment of SM backgrounds, we follow the analysis by ATLAS~\cite{ATLAS:2016cyf}. 
There are multiple sources of backgrounds. The dominant contribution arises from 
the DY process, where the muon pair is produced via $Z$/$\gamma^*$.
Other nonnegligible contributions arise from $t\bar{t}$, $Wt$ and $WW$ production,
while contributions from $WZ$ and $ZZ$ production are less significant.
As in the $tZ'$ case, we do not include backgrounds associated with nonprompt leptons.

The signal and background samples for the dimuon process are generated in a similar
way as in the previous subsection, except the treatment of additional jets.
In this case, we generate matrix elements of signal and backgrounds with up to 
two additional jets, followed by showering.
The LO $Z/\gamma^*$ (DY) cross section is normalized to 
the NNLO QCD+NLO EW one with LO photon-induced channel 
by the correction factor 1.27. The latter is obtained by FEWZ 3.1~\cite{Li:2012wna} 
in the dimuon-invariant mass range of $m_{\mu\mu} > 106$ GeV.
The LO $t\bar{t}$ and $Wt$ cross sections are normalized to the NNLO+NNLL ones
by the factors $1.84$~\cite{twiki} and $1.35$~\cite{Kidonakis:2010ux}, respectively. 
As for $WW$, $WZ$ and $ZZ$, the LO cross sections are normalized 
to the NNLO QCD ones by the factors $1.98$~\cite{Gehrmann:2014fva}, $2.07$~\cite{Grazzini:2016swo} and $1.74$~\cite{Cascioli:2014yka}, respectively.

Normalized distributions of the dimuon invariant mass $m_{\mu\mu}$ 
are given in Fig.~\ref{invmass} for the dimuon process in Case I, II and the backgrounds,
obtained by close-to-default cuts in MadGraph. 
The $p_T$ distributions of the leading and subleading muons are given in Fig.~\ref{ptdist}.
We apply two sets of cuts on signal and background events
as in the previous subsection.

\begin{table*}[t!]
\centering
\begin{tabular}{|c|c|c|c|c|c|c|c|c|c|c|}
\hline
\hline
Cuts                                    & Signal (Case I)    & $Z/\gamma^*$  & $t\bar{t}$    & $Wt$    & $WW$  & $WZ$     & $ZZ$ & Total BG         \\
\hline
\hline
\textbf{Pre-selection cuts }            & 38.65            & 19980        & 1785            & 166   & 212  & 128.44  &  74.82 & 22346            \\ 
&&&&&&&&\\
\textbf{Selection cuts}                 & 20.96            & 1677          & 163             & 16   & 24    & 0.22    &  0.02 & 1880          \\
\hline
\end{tabular}
\caption{Same as Table \ref{cut_table_zpt150} (cross sections in fb), but for the dimuon process in 
Case I ($m_{Z'}=150$ GeV).
}\label{cut_table_zp_150}
\end{table*}
\begin{table*}[t!]
\centering
\begin{tabular}{|c|c|c|c|c|c|c|c|c|c|c|}
\hline
\hline
Cuts                                 & Signal (Case II) & $Z/\gamma^*$  & $t\bar{t}$ & $Wt$    & $WW$   & $WZ$   & $ZZ$ & Total BG    \\
\hline
\hline
\textbf{Pre-selection cuts }           & 17.77           & 19980        & 1785       & 166     & 212  & 128.44  &  74.82 & 22346      \\ 
&&&&&&&&\\
\textbf{Selection cuts}                & 10.22           & 532          & 117        & 12      & 14   & 0.12     &  0.01  & 675       \\
\hline
\end{tabular}
\caption{Same as Table \ref{cut_table_zpt150}, but for the dimuon process in 
Case II ($m_{Z'}=200$ GeV).
}\label{cut_table_zp_200}
\end{table*}

\noindent\textbf{Pre-selection cuts}: 
The two muons in an event are required to have transverse momenta
$p_T^{\mu}> 50$ GeV, maximum pseudo-rapidity $\left|\eta\right|^\mu < 2.5$,
with minimum separation $\Delta R > 0.4$.

\noindent\textbf{Selection cuts}: 
Events are selected such that each event contains two oppositely-charged muons with
leading muon transverse momentum $p_T^{\mu_1}>$ 60 (75) GeV, and 
subleading muon $p_T^{\mu_2}>$ 55 (60) GeV in Case I (II).  
We impose an invariant-mass cut of $|m_{\mu\mu}-m_{Z'}|<15$ GeV on the two muons 
in both Case I and II.

The effects of the two sets of cuts on the signal and backgrounds are tabulated 
in Table~\ref{cut_table_zp_150} for Case I, and in Table~\ref{cut_table_zp_200} for Case II. 
From these tables, we see that the number of background events $B$
is significantly larger than signal events $S$ in both Case I and II, 
even after the selection cuts.
In this case, the signal significance of Eq.~\eqref{eq:signif} becomes 
$\mathcal{Z}\simeq S/\sqrt{B}$, which we use to estimate 
the discovery potential of the dimuon process.
We find that the $Z'$ in benchmark Case I (II) can be discovered in the dimuon process at $5\sigma$ with
integrated luminosity of $\mathcal{L}=110~(170)$ fb$^{-1}$.
We remark that in actual experimental searches the $Z'$ mass would be scanned 
over a certain range and the look-elsewhere effect would be included.
The latter effect will reduce the signal significance we estimated, 
pushing the integrated luminosities required for discovery to higher values.

\section{Discovery potential}
\label{const}

In this section, we first extend the results of the previous section to higher $Z'$ masses 
within the effective theory framework of the RH $tcZ'$ and $ccZ'$ couplings, 
then give the discovery potential of the $Z'$ in the $ tZ'$ and 
dimuon processes at the 14 TeV LHC.
We then reinterpret these model-independent results based on
the gauged $L_\mu-L_\tau$ model~\cite{Altmannshofer:2014cfa}.
We also discuss the sensitivity on the LH $tcZ'$ coupling, directly linked to
the $P_5'$ and $R_K$ anomalies.

\subsection{Model-independent results}
\label{effcoupling}

In the previous section, we studied the $tZ'$ and dimuon processes for $m_{Z'}=150$
and 200 GeV with benchmark values of the effective couplings $g_{ct}^R$ and $g_{cc}^R$. 
In this subsection, we extend the analysis to higher $Z'$ masses up to 700 GeV and
to arbitrary values of $g_{ct}^R$ and $g_{cc}^R$, and illustrate the $Z'$ discovery potential 
at the 14 TeV LHC with 300 and 3000 fb$^{-1}$ integrated luminosities. 

For $m_{Z'}=150$ and 200 GeV, we simply rescale the results of the previous section
by $|g_{ct}^R|$ and $|g_{cc}^R|$.
For higher $Z'$ masses from 300 to 700 GeV, in steps of 100 GeV, we follow the same
ways as the 200 GeV case for the generation of events and the application of cuts.
In particular, we adopt the same dimuon-invariant mass cut of $|m_{\mu\mu}-m_{Z'}|<15$ GeV.
We choose a $Z'$ width such that $\Gamma_{Z'}/m_{Z'} \lesssim 1$\% is satisfied 
for each mass. We assume $\mathcal{B}(Z'\to \mu^+\mu^-)= 1/3$.

We do not consider lower $Z'$ masses, as control of SM backgrounds 
becomes more difficult toward $m_{Z'} \sim m_Z$. 
We leave this case for future analysis.
We restrict the analysis for the $Z'$ mass up to 700 GeV, as the $S$ and $B$ for 
the $tZ'$ process, obtained from Eq.~\eqref{eq:signif} with 5$\sigma$, 
get smaller than $\mathcal{O}(1)$ beyond this mass.
On the other hand, for the dimuon process, the $S/B$ ratios become very low for masses
beyond 700 GeV, and proper understanding of systematic uncertainties would be needed.

The discovery reach for the effective couplings $g_{ct}^R$
and $g_{cc}^R$ are shown in the left and right panels of Fig.~\ref{eff}, respectively, 
for $150~{\rm GeV} \leq m_{Z'} \leq 700~{\rm GeV}$.
In the left panel, the upper red (lower blue) solid line represents the 5$\sigma$ discovery reach
for the $ tZ'$ process with 300 (3000) fb$^{-1}$ integrated luminosity, while
the corresponding dashed lines represent the 3$\sigma$ reach.
In the right panel, the discovery reaches for the dimuon process are similarly shown;
in this case, existing LHC results for dimuon resonance searches already constrain 
$g_{cc}^R$, as discussed in Sec.~\ref{dimuonc}, and the 95\% CL exclusion set by 
ATLAS \cite{ATLAS:2016cyf} (CMS \cite{CMS:2016abv}) with around 13 fb$^{-1}$ of 
13 TeV data is shown by the gray (semi-transparent blue) shaded region.

\begin{figure*}[t!]
  \centering
  \raisebox{0.4mm}{\includegraphics[width=.37\linewidth]{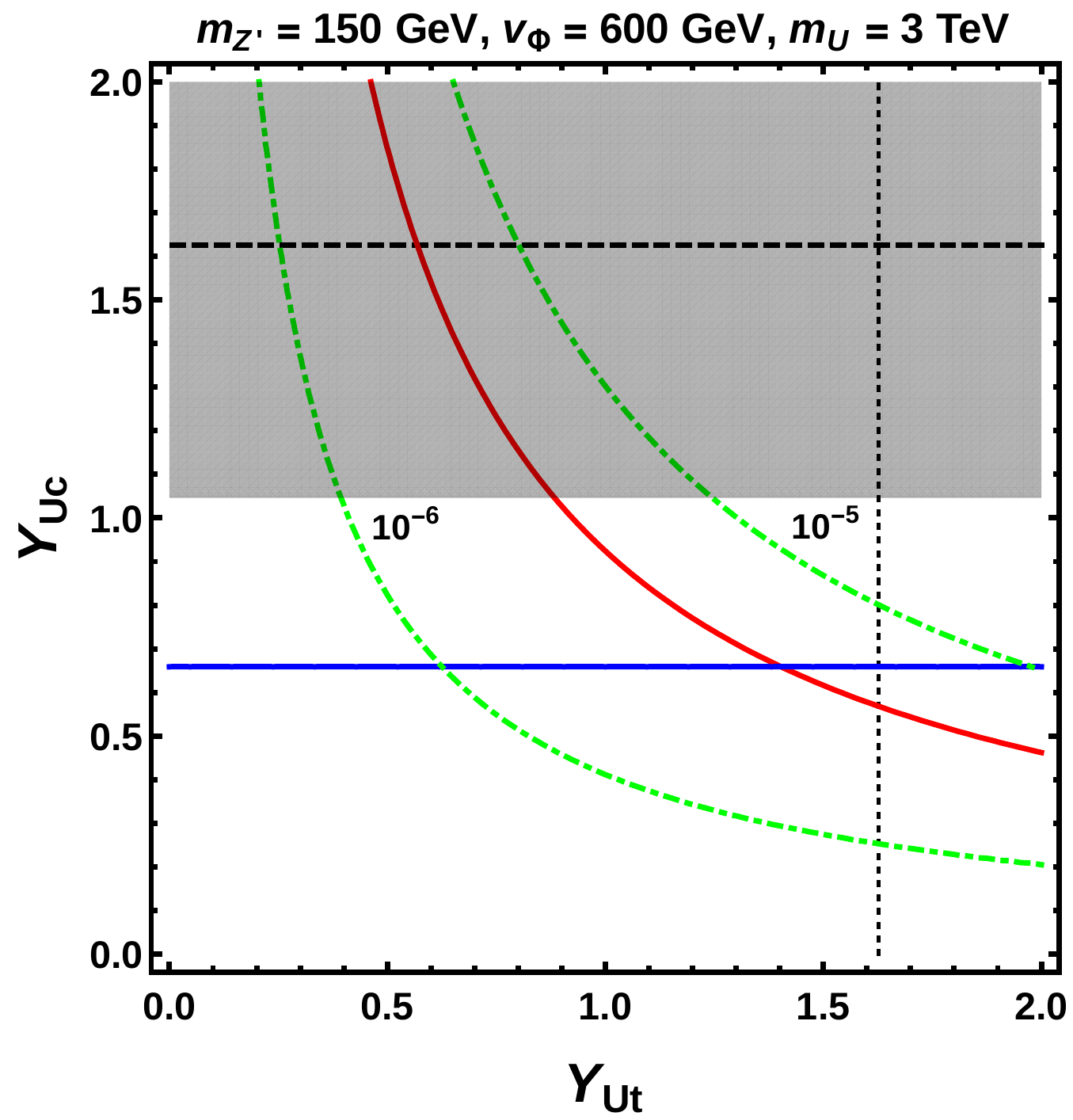}}
  \includegraphics[width=.564\linewidth]{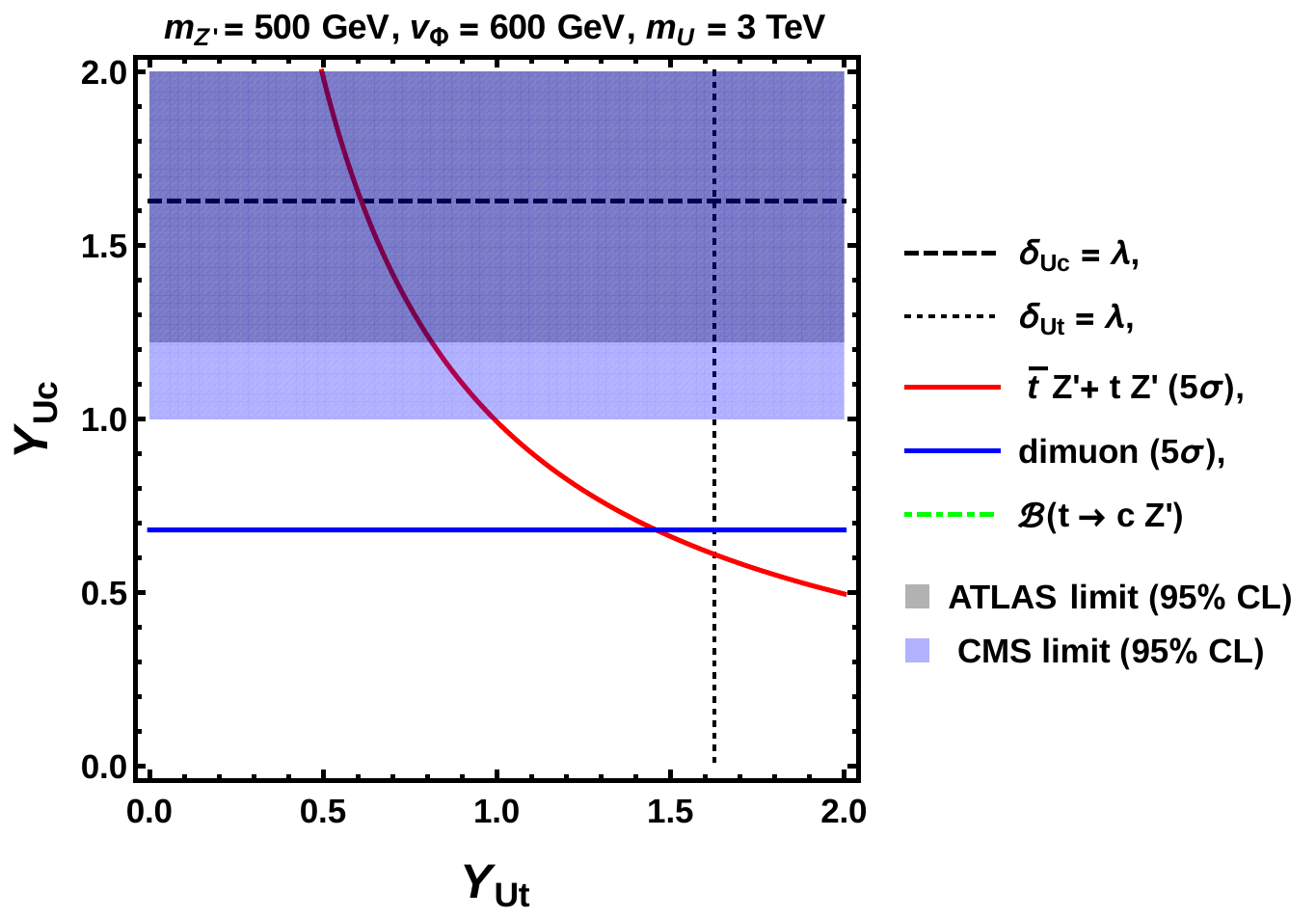}
\caption{
[Left] 5$\sigma$ discovery reach in $Y_{Ut}$--$Y_{Uc}$ plane for $m_{Z'}=150$ GeV 
with $v_\Phi = 600$ GeV ($\Gamma_{Z'}\simeq 0.74$ GeV) and $m_U=3$~TeV with 3000 fb$^{-1}$ data:
red solid line for the $tZ'$ process, and horizontal blue solid line for the dimuon process.
Green dash-dot lines are contours for $\mathcal{B}(t\to c Z')=10^{-6}$ and $10^{-5}$.
The gray shaded region is the 95\% CL exclusion by the ATLAS dimuon resonance 
search \cite{ATLAS:2016cyf}.
The mixing parameter $\delta_{Ut}$ ($\delta_{Uc}$) exceeds $\lambda\simeq 0.23$
beyond the vertical dotted (horizontal dashed) line. 
[Right] Same as left panel, but for $m_{Z'}=500$ GeV ($\Gamma_{Z'}\simeq 27$ GeV).
The CMS \cite{CMS:2016abv} 95\% CL exclusion, shown by the semi-transparent 
blue shaded region, is overlaid on the gray shaded ATLAS exclusion and gives
stronger constraint.
}\label{fig:yukawa}
\end{figure*}


We see that, at the 14 TeV LHC with 300 (3000) fb$^{-1}$ data, the $ tZ'$ process 
can be discovered for $|g_{ct}^R|=0.025$ up to $m_{Z'}\simeq 490~(700)$ GeV; the dimuon
process can be discovered for $|g_{cc}^R|=0.01$ up to $m_{Z'}\simeq 460~(650)$ GeV.
We also read the discovery reach for representative $Z'$ mass values: 
$|g_{ct}^R|\gtrsim 0.0086~(0.0047)$ and $|g_{cc}^R|\gtrsim 0.0039~(0.0022)$
for $m_{Z'}=150$ GeV;
$|g_{ct}^R|\gtrsim 0.026~(0.013)$ and $|g_{cc}^R|\gtrsim 0.011~(0.0063)$ for $m_{Z'}=500$ GeV,
with 300 (3000) fb$^{-1}$ data.
The dimuon process can probe smaller $Z'$ couplings than the $tZ'$ process, 
but these two couplings are independent in general.

The results in Fig.~\ref{eff} are model independent,
except for the assumptions of narrow $Z'$ width and $\mathcal{B}(Z'\to \mu^+ \mu^-)=1/3$, 
which are motivated by the gauged $L_\mu -L_\tau$ model.
For arbitrary $\mathcal{B}(Z'\to \mu^+ \mu^-)$,
the discovery reach can be obtained from Fig.~\ref{eff} by simply replacing
\begin{align}
|g_{ct}^R| \to |g_{ct}^R|\sqrt{3\times\mathcal{B}(Z'\to\mu^+\mu^-)},
\end{align}
with similar replacement for $|g_{cc}^R|$.
We remark that the same results apply to the LH coupling $g_{ct}^L$ ($g_{cc}^L$) 
if the RH coupling $g_{ct}^R$ ($g_{cc}^R$) is set to zero.


%
\subsection{Interpretation in the gauged $L_\mu-L_\tau$ model}
\label{glmutau}

Both the $tZ'$ and dimuon processes can probe the effective $Z'$ couplings
implied by the gauged $L_\mu -L_\tau$ model \cite{Altmannshofer:2014cfa}.
In this subsection, we reinterpret the model-independent results of the previous subsection 
within the gauged $L_\mu -L_\tau$ model\footnote{
In the gauged $L_\mu-L_\tau$ model, a nonzero $g_{ct}^R$ is accompanied with 
a nonzero $g_{tt}^R$, leading to the $gg \to t\bar{t}Z'$ process.
The latter could contribute to the signal region of the $tZ'$ process despite the veto 
on extra jets. 
We, however, found that such a contribution is smaller than 1\% for $|g_{ct}^R| \sim |g_{tt}^R|$.
We ignore the effects from the $t\bar tZ'$ production in the following analysis.}
through the expressions for $g_{ct}^R$ and
$g_{cc}^R$ in Eqs.~\eqref{eq:gctR} and \eqref{eq:gccR}, and discuss the discovery
potential at the 14 TeV LHC with 3000 fb$^{-1}$ data.

\begin{figure*}[t!]
  \centering
  \includegraphics[width=.40\linewidth]{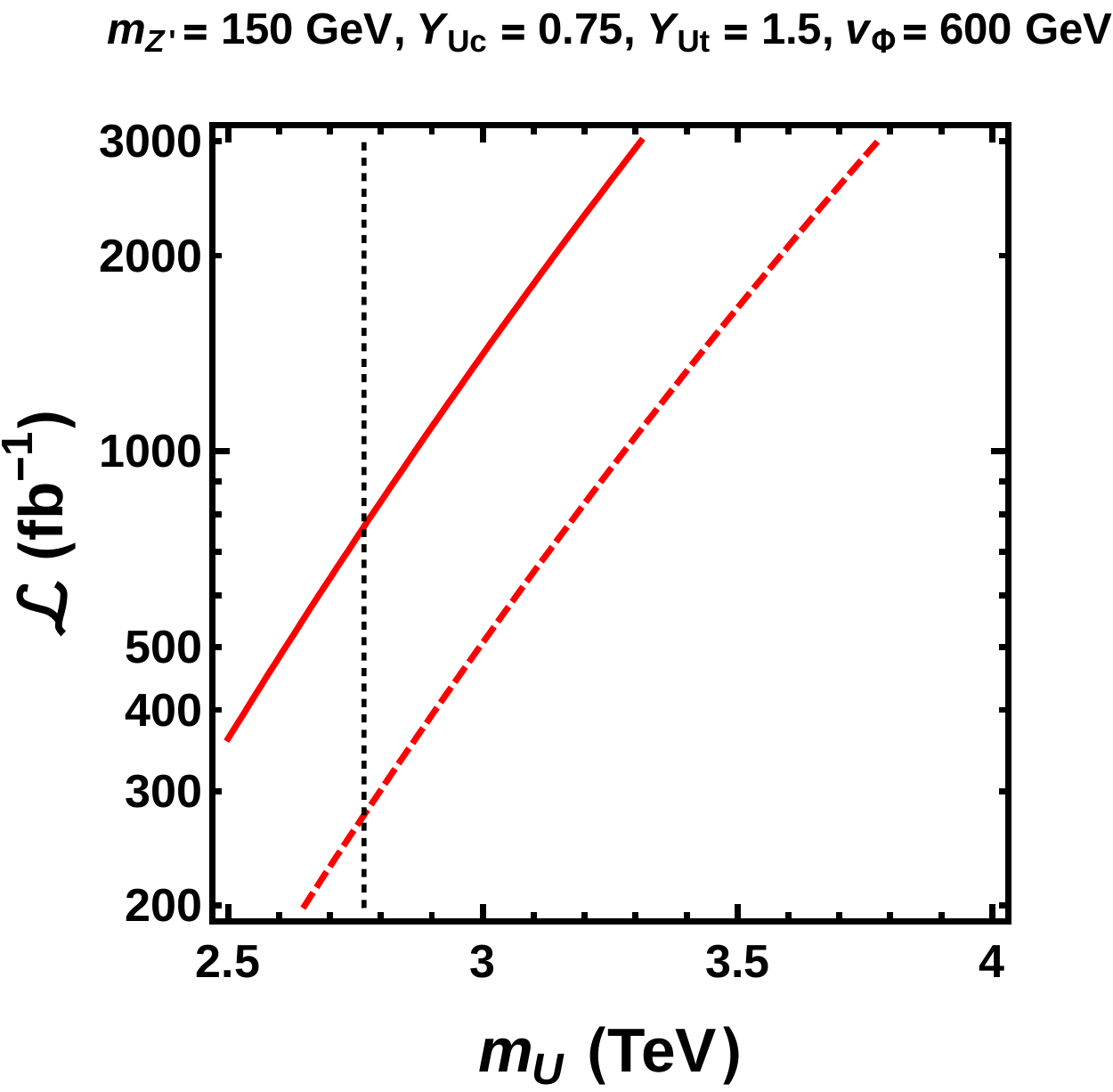}
  \raisebox{-0.4mm}{\includegraphics[width=.59\linewidth]{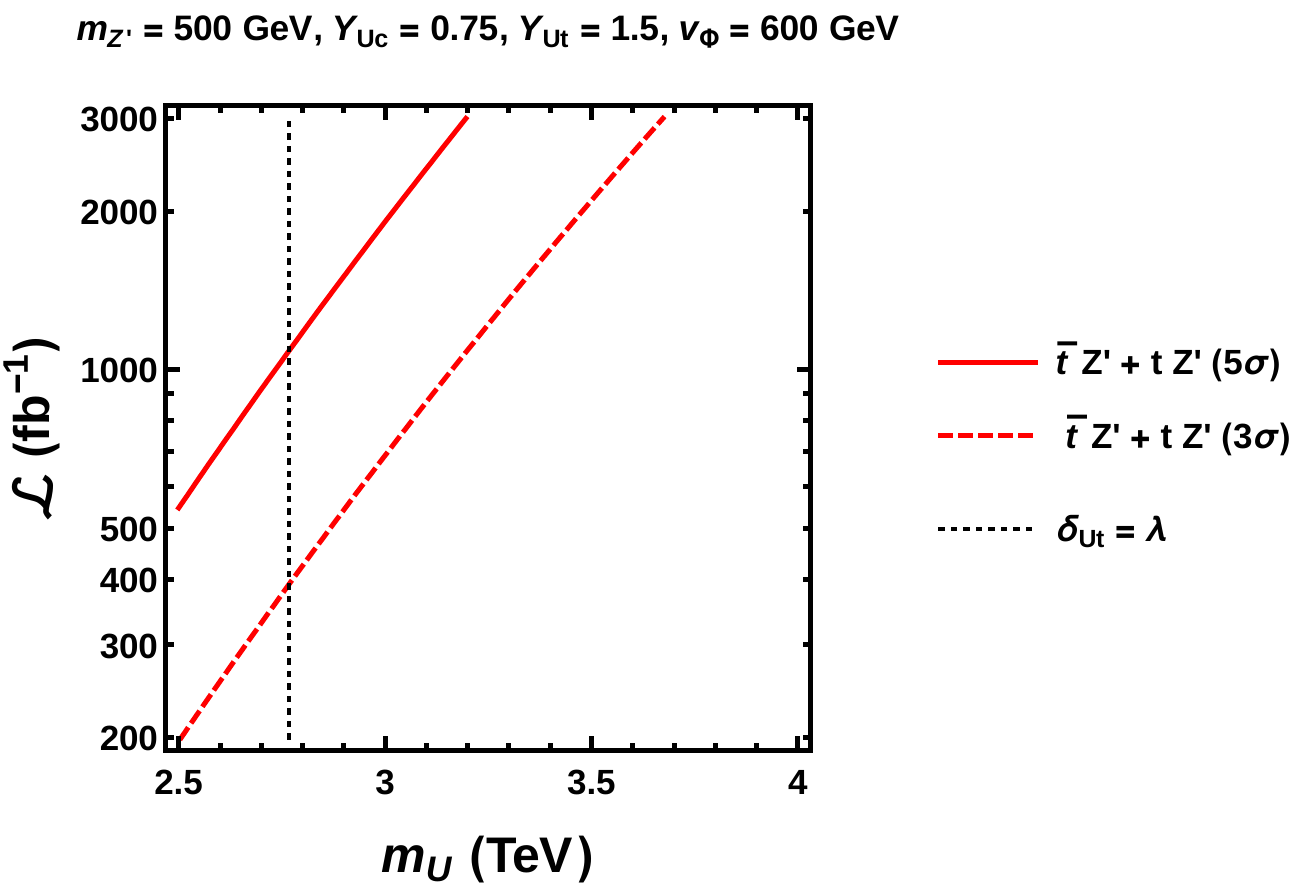}} 
\caption{
Integrated luminosities needed for 5$\sigma$ discovery (red solid lines) of 
the $tZ'$ process at the 14 TeV LHC as a function of vector-like quark 
mass $m_U$ with $Y_{Ut} = 1.5$, $Y_{Uc} = 0.75$ and $v_\Phi = 600$ GeV, for 
$m_{Z'}=$ 150 GeV [left] and 500 GeV [right]. 
The 3$\sigma$ reach is given by red dashed lines, while
vertical dotted lines indicate the $m_U$ value below which the mixing parameter
$\delta_{Ut}$ exceeds $\lambda\simeq 0.23$. 
}
\label{fig:mU-tZp}
\end{figure*}

From Eq.~\eqref{eq:gctR-range}, one can observe that a smaller $v_\Phi$ is better
probed for fixed $m_{Z'}$ and mixing parameters $\delta_{Ut}$ and $\delta_{Uc}$.
Applying the 5$\sigma$ discovery reach of Fig.~\ref{eff}, we find that
the $tZ'$ process can be discovered for $m_{Z'}=150~(500)$ GeV with 
$\delta_{Ut} = \delta_{Uc} =\lambda \simeq 0.23$ if $v_\Phi \lesssim 1.7~(2.0)$ TeV;
the dimuon process can be discovered for the same parameters 
if $v_\Phi \lesssim 3.6~(4.2)$ TeV.
In general, if the two Yukawa couplings have a same value, i.e.
$\delta_{Uc}/\delta_{Ut}=Y_{Uc}/Y_{Ut}=1$, we find the dimuon process 
to have better discovery potential.
If the Yukawa couplings are hierarchical such that $Y_{Uc}/Y_{Ut} = \lambda$,
the discovery reach of the $ tZ'$ process becomes $v_\Phi \lesssim 390~(470)$ GeV,
which is better than the dimuon process $v_\Phi \lesssim 190~(220)$ GeV
for $m_{Z'}=150~(500)$ GeV with $\delta_{Ut} = \lambda$.
These $v_\Phi$ values are, however, already excluded by the neutrino trident production
[See Eq.~(\ref{eq:vphi-range})].
Taking a milder hierarchy such that $Y_{Uc}/Y_{Ut} \simeq 0.47~(0.48)$, we find
a comparable discovery reach between the two processes: $v_\Phi \lesssim 790~(990)$ GeV
for $m_{Z'}=150~(500)$ GeV with $\delta_{Ut} = \lambda$.
In this case, the two processes can probe the parameter region allowed by 
the neutrino trident production. For a slightly smaller $Y_{Uc}/Y_{Ut}$, the $ tZ'$ process
can have better discovery potential than the dimuon process with
neutrino trident production constraint satisfied.

The mixing parameters $\delta_{Ut}$ and $\delta_{Uc}$, defined in Eq.~(\ref{eq:deltaUq}),
depend on $Y_{Ut}$, $Y_{Uc}$, $m_U$ as well as $v_\Phi (=m_{Z'}/g')$.
In Fig.~\ref{fig:yukawa}, we show the impact of the Yukawa couplings on the discovery
of the $Z'$ by taking $v_\Phi=600$ GeV, close to the lower end of Eq.~(\ref{eq:vphi-range}),
to maximize the discovery reach.
We also fix $m_U=3$ TeV, but different choices of $m_U$ will just give rescaled figures.

In the left panel of Fig.~\ref{fig:yukawa}, the discovery reach is given in 
the $Y_{Ut}$--$Y_{Uc}$ plane for $m_{Z'}=150$ GeV.
The red and horizontal blue solid lines represent the discovery reach for 
the $tZ'$ and dimuon processes, respectively.
We only consider the parameter region where the mixing parameters satisfy 
$|\delta_{Ut}|, |\delta_{Uc}| \leq \lambda$, shown by the vertical dotted
and horizontal dashed lines, respectively.
The gray shaded region represents the 95\% CL exclusion by the dimuon resonance search of
ATLAS \cite{ATLAS:2016cyf}. The latter can already
probe the parameter region that satisfies $|\delta_{Uc}| \leq \lambda$.
The dimuon process can be discovered for $Y_{Uc}\gtrsim 0.7$, and generally has 
a larger discovery zone than the $ t Z'$ process, in particular for small $Y_{Ut}$.
Interestingly, there is an overlap of discovery zones of the two processes
for $Y_{Ut}\gtrsim 0.9$ and $Y_{Uc}\gtrsim 0.7$, 
and discovery might be possible for both processes.
This might be useful to probe the flavor structure of the model.
As the $Z'$ is lighter than the top quark, $t\to cZ'$ may happen.
The green dash-dot contours are plotted for $\mathcal{B}(t\to cZ')=10^{-6}$ and $10^{-5}$.
One can see that the $ t Z'$ process can probe the region where 
$\mathcal{B}(t\to cZ')<10^{-5}$.

In the right panel of Fig.~\ref{fig:yukawa}, a similar plot is shown for $m_{Z'}=500$ GeV.
We again take $v_\Phi = 600$ GeV, which gives the $Z'$ width of $\Gamma_{Z'}\simeq 27$
GeV. This is rather large and the dimuon-invariant mass distribution would spread 
outside the invariant mass cut $|m_{\mu\mu}-m_{Z'}|<15$~GeV, applied in our collider
study of the last subsection with the narrow width assumption.
Hence, the discovery reaches shown in the last subsection do not apply.
In order to evaluate the discovery potential in this case, 
we regenerated the signal events for the case of $m_{Z'}=500$ GeV with
$\Gamma_{Z'}\simeq 27$ GeV and redid the cut-based analysis with the same
cuts as the narrow width case, but relaxing the invariant mass cut to
$|m_{\mu\mu}-m_{Z'}|<55$ GeV. We then obtain the model-independent
discovery reach:
\begin{align}
|g_{ct}^R|\gtrsim 0.016,\quad |g_{cc}^R|\gtrsim 0.0077, \quad (\Gamma_{Z'}= 27~{\rm GeV})
\end{align}
at $m_{Z'}=500$ GeV for 3000 fb$^{-1}$ data.
The result gets slightly worse due to increased number of SM background events.
With these results, we plot in the right panel of Fig.~\ref{fig:yukawa} 
the discovery reach for $m_{Z'}=500$ GeV.
The qualitative feature is similar to the $m_{Z'}=150$ GeV case,
but the ATLAS constraint is now weaker than the CMS~\cite{CMS:2016abv} 95\% CL limit 
on dimuon resonance search, as illustrated by the gray shaded region being overlaid by 
the semi-transparent blue shaded region.

\begin{figure*}[t!]
  \centering
  \includegraphics[width=.38\linewidth]{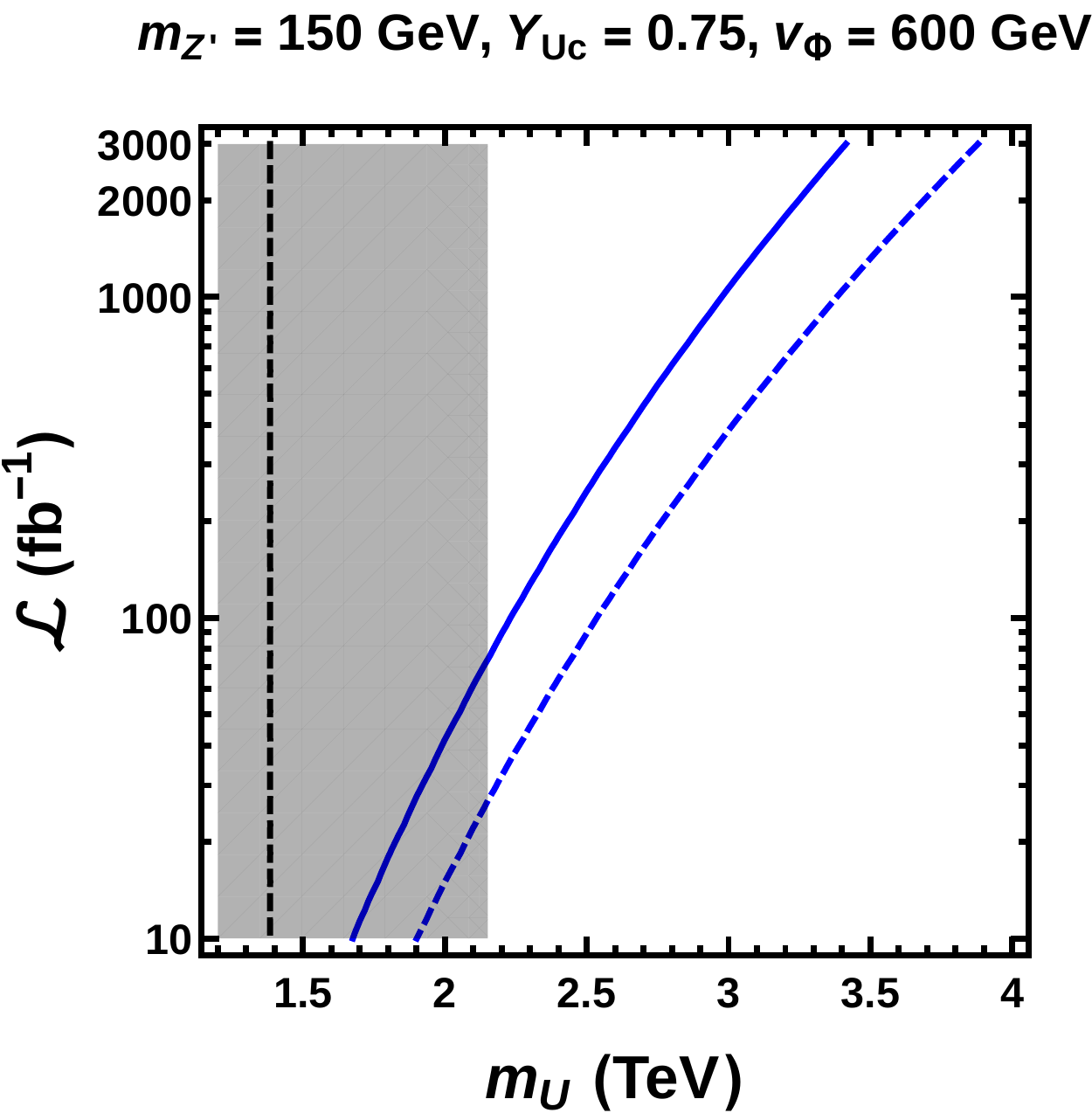}
  \raisebox{0mm}{\includegraphics[width=.56\linewidth]{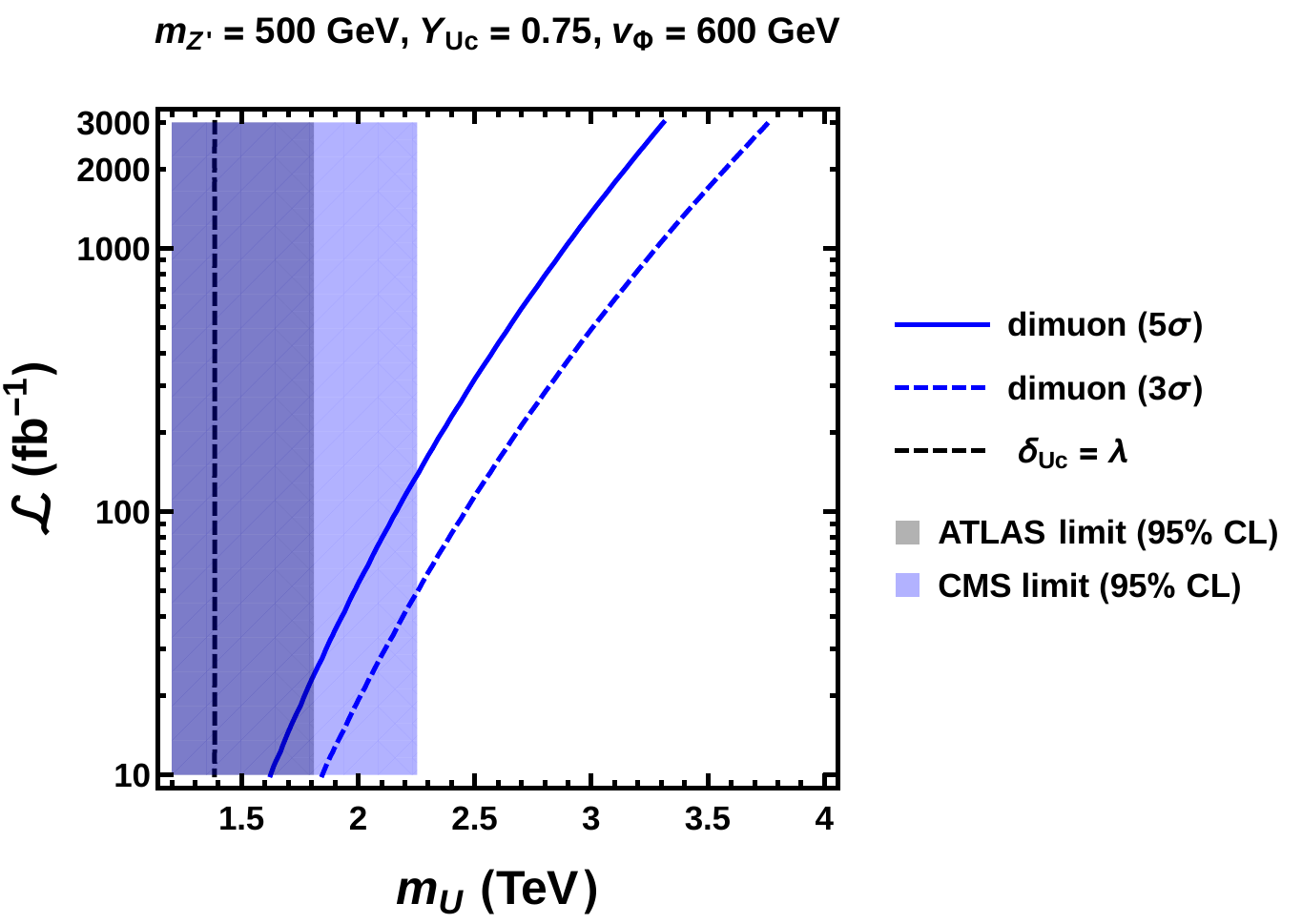}}
\caption{Same as Fig.~\ref{fig:mU-tZp}, but for the dimuon process with 
gray (semi-transparent blue) shaded region showing the 95\% CL exclusion by the 
dimuon resonance search of ATLAS \cite{ATLAS:2016cyf} (CMS \cite{CMS:2016abv}).
}\label{fig:mU-dimuon}
\end{figure*}

Fixing the Yukawa couplings, we can see the {\it indirect} discovery reach for the vector-like
quark mass scale $m_U$. For illustration, we take a hierarchical pattern of the Yukawa 
couplings $Y_{Ut} = 1.5$ and $Y_{Uc} = 0.75$ with $v_\Phi = 600$ GeV.
In Fig.~\ref{fig:mU-tZp}, integrated luminosities required for the discovery of the $ t Z'$
process are shown by red solid lines as a function of $m_U$ for $m_{Z'}=$ 150 GeV [left] 
and 500 GeV [right]. The red dashed lines are for the 3$\sigma$ reaches.
The vertical dotted lines mark the minimum value of $m_U$ satisfying 
the small mixing condition $|\delta_{Ut}| \leq \lambda$. 
With 3000 fb$^{-1}$ data, discovery of $m_U \simeq 3$~TeV is possible 
for both $m_{Z'}$ cases.

Similar plots for the dimuon process are given in Fig.~\ref{fig:mU-dimuon} 
for $m_{Z'}=$ 150 GeV [left] and 500 GeV [right].
The gray (semi-transparent blue) shaded region shows the 95\% CL exclusion by 
the dimuon resonance search of ATLAS \cite{ATLAS:2016cyf} (CMS \cite{CMS:2016abv}), 
as in Fig.~\ref{fig:yukawa}.
With 3000 fb$^{-1}$ data, discovery is possible for $m_U \gtrsim 3$ TeV in both $m_{Z'}$ cases.
The dimuon resonance search limits push up $m_U$, such that discovery 
is possible for the $m_{Z'}=$ 150 (500) GeV case after $\sim$80 (110) fb$^{-1}$ data
is accumulated at the 14 TeV LHC,
while 3$\sigma$ evidence can be made with $\sim$30 (50) fb$^{-1}$ data. 
This means discovery could be made with LHC Run 2 data,
where experiments can easily change between 13 and 14 TeV collision energies.

%
Note that the ATLAS and CMS 95\% CL limits assume narrow $Z'$ width,
while our discovery reach for $m_{Z'}=500$ GeV is estimated with 
rather large width ($\Gamma_{Z'}\simeq 27$ GeV).
Note also that CMS gives stronger limit for $m_{Z'}=500$ GeV, 
in part due to the observed limit being better than expected by  $\sim 1\sigma$ \cite{CMS:2016abv}, 
while our discovery reach was estimated by inclusion of ATLAS-based detector effects.
We have not taken into account systematic uncertainties and backgrounds associated with 
nonprompt leptons. 
These would lead to uncertainties in the integrated luminosity for discovery quoted above.

\subsection{Sensitivity for LH $tcZ'$ coupling motivated by $P_5'$ and $R_K$ anomalies}
\label{senlh}

So far we concentrated on the RH $tcZ'$ coupling, which is inspired by, but 
not directly linked with, the $P_5'$ and $R_K$ anomalies. 
Let us now discuss the LH $tcZ'$ coupling that is directly linked to these anomalies.

The LH $tcZ'$ and $bsZ'$ couplings are related by the SU(2)$_L$ relation of 
Eq.~\eqref{eq:Yukawa-SU(2)}: $g_{ct}^L \simeq V_{cs}V_{tb}^* g_{sb}^L +V_{cb}V_{tb}^* g_{bb}^L
\sim g_{sb}^L +\lambda^2 g_{bb}^L$, where CKM suppressed terms are neglected except 
the $g_{bb}^L$ term.
Using Eq.~\eqref{eq:C9}, one can express the first term as 
$g_{sb}^L =m_{Z'}v_\Phi\Delta C_9^\mu$. The upper and lower limits for $v_\Phi$ 
in Eq.~\eqref{eq:vphi-range} then leads to
\begin{align}
 0.7\times 10^{-4}&\bigg(\frac{m_{Z'}}{150~{\rm GeV}}\bigg)\left( 
  \frac{\left|\Delta C^\mu_9\right|}{(34~{\rm TeV})^{-2}} \right) \nn\\
  &\lesssim |g_{sb}^L |
  \lesssim 0.7\times 10^{-3}\bigg(\frac{m_{Z'}}{150~{\rm GeV}}\bigg). \label{eq:gsbL-range}
\end{align}
Here, the best-fit value of $\Delta C^\mu_9\simeq -(34~{\rm TeV})^{-2}$ from a recent 
global analysis \cite{Descotes-Genon:2015uva} is used in the lower limit, while its dependence 
is canceled out in the upper limit. 
The latter is set by the $B_s$ mixing constraint with the $Z'$ effect allowed within 15\%.
The $g_{bb}^L$ term can be as large as the $g_{sb}^L$ term if the Yukawa couplings are hierarchical, 
such that $|g_{bb}^L/g_{sb}^L|= |Y_{Qb}/Y_{Qs}| \sim \lambda^{-2}$, which is indeed 
advocated in Ref.~\cite{Altmannshofer:2014cfa} as a viable solution for the $P_5'$ anomaly.
However, the relative sign of the two terms are opposite because $\Delta C^\mu_9 < 0$
implies a negative $g_{sb}^L$ while $g_{bb}^L$ is positive by definition [See Eq.~(\ref{effcoup})].
Hence a large $g_{bb}^L$ tends to suppress $g_{ct}^L$.

We thus conclude that $|g_{ct}^L|$ can not be larger than $|g_{sb}^L|$.
The latter is constrained by Eq.~(\ref{eq:gsbL-range}).
We then obtain the upper limits on the LH $tcZ'$ coupling:
\begin{align}
 \left|g_{ct}^L\right|_{\rm max}\sim 
 \begin{cases}
   1\times 10^{-3} & (m_{Z'}=150~{\rm GeV}), \\
   2\times 10^{-3} & (m_{Z'}=500~{\rm GeV}), \label{eq:LHct}\\
   3\times 10^{-3} & (m_{Z'}=700~{\rm GeV}).
 \end{cases}
\end{align}
If we set $g_{ct}^R=0$, we can directly apply the discovery reach of Fig.~\ref{eff} [left] 
to the LH coupling $g_{ct}^L$.
We find that, for the $P_5'$ and $R_K$ motivated case, the maximally allowed 
values of $|g_{ct}^L|$ are beyond (i.e. smaller than) the discovery reach with 
3000 fb$^{-1}$ data, by a factor of 5--10.
One cannot even attain 3$\sigma$ evidence for the $|g_{ct}^L|$ values given in Eq.~(\ref{eq:LHct}).

We remark that we have estimated the signal $ tZ'$ events at LO and have not taken 
into account QCD corrections, which may enhance the number of signal events.
Moreover, the discovery reach might be improved by combining ATLAS and CMS data.

We note in passing that the LH $ccZ'$ coupling is also related to the LH $bsZ'$ coupling, 
but in a more complicated way:
 $g_{cc}^L \simeq 2{\rm Re}(V_{cs}V_{cb}^* g_{sb}^L) +|V_{cs}|^2g_{ss}^L+|V_{cb}|^2g_{bb}^L$, 
where terms containing the $d$ quark are neglected with 
the choice of $Y_{Qd}\simeq 0$ for the $K$ and $B_d$ meson mixing constraints.
Choosing different Yukawa coupling hierarchies with $m_{Z'}=150$ GeV,
we find the following upper limits on $|g_{cc}^L|$ from the $B_s$ mixing constraint 
on $v_\Phi$ and the small mixing conditions $| \delta_{Qq} | \leq \lambda$
 [$q=s,c,b,t$, defined as in Eq.~\eqref{eq:deltaUq}]:
$|g_{cc}^L|\lesssim 6\times 10^{-4}$ for $Y_{Qb}=1$, $Y_{Qs}=-1$ and $m_Q =24$ TeV;
$|g_{cc}^L|\lesssim 3\times 10^{-7}$ for $Y_{Qb}=1$, $Y_{Qs}=-\lambda^2$ and $m_Q =5.4$ TeV;
$|g_{cc}^L|\lesssim 4\times 10^{-3}$ for $Y_{Qb}=\lambda^2$, $Y_{Qs}=-1$ and $m_Q =5.4$ TeV,
where the values for $m_Q$ are chosen such that the best-fit value of 
$\Delta C^\mu_9\simeq -(34~{\rm TeV})^{-2}$ is realized.
From the right panel of Fig.~\ref{eff}, we read the discovery reach of the dimuon process as 
$|g_{cc}^L|\gtrsim 2.3\times 10^{-3}$ for $m_{Z'}=150$ GeV with 3000 fb$^{-1}$ data.
Interestingly, the third case with a skewed Yukawa hierarchy $|Y_{Qb}/Y_{Qs}|=\lambda^2$
could be discovered, as long as $v_\Phi \gtrsim 1$ TeV.

\section{Summary and discussion}\label{summary}

The $P_5'$ and $R_K$ anomalies in $B\to K^{(*)}$ transitions may indicate the existence
of a new $Z'$ boson with FCNC couplings. In this paper, we studied the LHC signatures of 
the RH $tcZ'$ coupling ($g_{ct}^R$) that is inspired by, but not directly linked to, 
the $B\to K^{(*)}$ anomalies. We first examined the 
$tcZ'$-induced process $cg\to tZ' \to b\nu_\ell\ell^+\mu^+\mu^-$ ($tZ'$ process) 
and its conjugate process ($\bar t Z'$ process) at the 14 TeV LHC within the effective theory framework.
We then discussed the implications in a specific $Z'$ model, namely the 
gauged $L_\mu-L_\tau$ model of Ref.~\cite{Altmannshofer:2014cfa}.
In this model, the RH $tcZ'$ coupling is induced by mixings of the SU(2)$_L$-singlet 
vector-like quark $U$ with the top and charm quarks, which also induce the 
flavor-conserving $ccZ'$ coupling. 
We thus also considered the $c\bar c\to Z' \to \mu^+\mu^-$ (dimuon process) at the LHC. 
We performed a collider study taking into account detector effects and 
major SM background processes. 

We find that the $\bar tZ'$ process has a better chance for discovery than the $tZ'$
process because of smaller backgrounds, and the combination of the two processes,
which we also call the $t Z'$ process collectively, can further enhance discovery potential.
The $tZ'$ process can be discovered with
3000 fb$^{-1}$ data for the $Z'$ masses in 150--700 GeV,
with $|g_{ct}^R|= \mathcal{O}(0.01)$ and $\mathcal{B}(Z'\to\mu^+\mu^-)=1/3$, 
e.g. $|g_{ct}^R|\gtrsim$ 0.0047 (0.013) for $m_{Z'}=$ 150 (500) GeV. 
Reinterpreted within the gauged $L_\mu -L_\tau$ model, these results imply 
that one can discover the $Z'$ if the mixing parameters of the vector-like quark $U$ with 
the top and charm quarks, $\delta_{Ut}$ and $\delta_{Uc}$, are $\mathcal{O}(0.1)$
and the VEV of the exotic Higgs is not too large, i.e. $v_\Phi \lesssim$ 2 TeV.
In the model, the $tZ'$ and dimuon processes are correlated,
with the dimuon process having better discovery potential if $|\delta_{Ut}| \sim |\delta_{Uc}|$,
starting with LHC Run 2 data.
But if the mixings are hierarchical, such that $|\delta_{Uc}/\delta_{Ut}|\lesssim 0.4$, 
the $tZ'$ process would have better discovery potential.
However, $g_{ct}^R$ tends to be suppressed in this case, and discovery is not 
possible at the HL-LHC if $|\delta_{Uc}/\delta_{Ut}|\lesssim \lambda \simeq 0.23$ with 
$|\delta_{Ut}| \leq \lambda$ for $v_\Phi$ values allowed by the neutrino trident production.
We illustrated the discovery zones in the model imposing the existing ATLAS and CMS 
dimuon resonance search constraints, and showed that there exist interesting
parameter regions where both the $tZ'$ and dimuon processes can be discovered. 
If this is the case, the simultaneous measurement of the two processes
can uncover the flavor structure of the model.

We also discussed the sensitivity for the LH $tcZ'$ coupling $g_{ct}^L$ that is directly 
linked to the $B\to K^{(*)}$ anomalies. We first identified the range of the LH $bsZ'$ coupling 
$g_{sb}^L$ favored by the $b\to s\ell^+\ell^-$ transition data, and then obtained the upper limits 
on $|g_{ct}^L|$ using SU(2)$_L$ symmetry.
We find that the $|g_{ct}^L|$ values implied by the $B\to K^{(*)}$ anomalies are 
beyond the discovery reach of the $tZ'$ process at the HL-LHC.
However, the sensitivity might be improved by inclusion of QCD
corrections to the signal cross section, and/or by combining ATLAS and CMS data.

The gauged $L_\mu -L_\tau$ model further implies flavor-conserving $ttZ'$ couplings,
which lead to the $pp\to t\bar tZ'$ production process at the LHC.
This process may provide not only another discovery channel of the $Z'$, but also 
useful information on the flavor structure of the model.
In particular, the three production modes, namely $tZ'$, dimuon and $t\bar tZ'$ 
processes can be correlated by the dependence on 
the two Yukawa couplings $Y_{Ut}$ and $Y_{Uc}$.
We note that the $ccZ'$ couplings can be also probed through the $cg\to cZ'$ process,
if one has efficient charm tagging.
These will be studied elsewhere.

In this paper, we focused on collider signatures of the $Z'$ couplings to the top and 
charm quarks, but discovery of the $Z'$ may also come from the couplings to
the down-type quark sector. In particular, the gauged $L_\mu -L_\tau$ model
predicts a nonzero LH $bbZ'$ coupling $g_{bb}^L$ if the LH $bsZ'$ coupling exists.
The $bbZ'$ coupling induces the process $b\bar b \to Z' \to \mu^+\mu^-$
and can be searched in the similar way as the $ccZ'$ coupling at the LHC.
Taking for illustration $m_{Z'}=200$ GeV, $v_\Phi=1.5$ TeV, $Y_{Qb}=1$, $Y_{Qs}=-\lambda^2$
and $m_Q=24$ TeV, giving $\Delta C_9^\mu = -(34~{\rm TeV})^{-2}$ for the $B\to K^{(*)}$
anomalies, we find $g_{bb}^L\simeq 5\times 10^{-3}$ and the induced $Z'$ production
cross section of $\sigma(pp\to Z')\simeq 30$ fb at the 14 TeV LHC by MadGraph.
Multiplying $\mathcal{B}(Z'\to \mu^+\mu^-)\simeq 1/3$ and assuming similar cuts and 
detector effects as in the $ccZ'$-induced dimuon process, we obtain the cross section of
$\sigma(pp\to Z' \to \mu^+\mu^-)\simeq 4$ fb with the event selection cuts.
Utilizing the background cross sections for the dimuon process in Table \ref{cut_table_zp_200},
we then find that such a $Z'$ can be discovered with $\sim 1000$ fb$^{-1}$ integrated
luminosity. The $b\bar b\to Z'$ production process has also been studied in other $Z'$ models 
constructed for the $B\to K^{(*)}$ anomalies~\cite{Boucenna:2016qad,Ko:2017quv}.


We emphasize that the RH $tcZ'$ coupling 
cannot be constrained well by $B$ and $K$ physics, 
but is on similar footing as the current $B\to K^{(*)}$ anomalies. 
In particular, the coupling may exist even if the $P_5'$ and $R_K$ anomalies evaporate in the future. 
Hence, it is important to explore the RH $tcZ'$ coupling regardless of 
the fate of the $B\to K^{(*)}$ anomalies, with potential of
discovering a new $Z'$ gauge boson as a dimuon resonance
with weaker and FCNC quark couplings.
Our study therefore illustrates a unique role of {\it top} physics in the flavor program.
If discovery is made at the LHC, one would then need to probe the 
handedness of the coupling via angular distributions,
while $c\bar c \to Z'$ discovery (and maybe also $cg \to cZ'$ and $t\bar tZ'$)
would provide complementary information, opening up a rich program.

\begin{acknowledgments}
We thank Y. Chao for discussions.
WSH is supported by grants MOST 104-2112-M-002-017-MY2, MOST 105-2112-M-002-018 and NTU 105R8965, 
MK is supported by NTU-105R104022 and NTU-G029927, 
and TM is supported by MOST-104-2112-M-002-017-MY2.
\end{acknowledgments}

\end{document}